\documentclass[preprintnumbers,nofootinbib,noshowpacs,eqsecnum,twocolumn,prd]{revtex4}

\usepackage{graphicx}
\usepackage{amsmath,amsthm,amssymb}
\usepackage{mathrsfs,bbm}
\usepackage[colorlinks=true,linkcolor=blue,urlcolor=blue,citecolor=blue]{hyperref}
\usepackage{enumitem}

\makeatletter       

\renewcommand{\p@subsection}{}

\makeatother

\newtheorem{theorem}{Theorem}

\newcommand*{\smgroup}{\mbox{$SU(3)_C \times SU(2)_L \times U(1)_Y$} }
\newcommand*{\eweakgroup}{\mbox{$SU(2)_L \times U(1)_Y$} }
\newcommand*{\emgroup}{\mbox{$U(1)_{em}$} }

\newcommand*{\CP}{\mbox{\it CP} }
\newcommand*{\unitmatrix}{\mathbbm{1}}
\newcommand*{\realnums}{\mathbbm{R}}
\newcommand*{\complexnums}{\mathbbm{C}}

\newcommand*{\abs}[1]{\left\lvert {#1} \right\rvert} 
\newcommand*{\twomat}[1]{\underline{#1}}             
\newcommand*{\tvec}[1]{\boldsymbol{#1}}              
\newcommand*{\fvec}[1]{\boldsymbol{\tilde{#1}}}      
\newcommand*{\fmat}[1]{\tilde{#1}}                   
\newcommand*{\trans}{\mathrm{T}}                     
\newcommand*{\oind}[1]{{\{#1\}}}                     
\newcommand*{\by}{\!\times\!}                        

\DeclareMathOperator{\trace}{tr}
\DeclareMathOperator{\mRe}{Re}
\DeclareMathOperator{\mIm}{Im}

\begin{document}

\preprint{HD-THEP-06-07}

\title{Stability and Symmetry Breaking in the General\\
       Two-Higgs-Doublet Model}

\author{M. Maniatis}
    \email[E-mail: ]{MManiatis@ubiobio.cl}
\author{A. von Manteuffel}
    \email[E-mail: ]{manteuffel@ur.de}
\author{O. Nachtmann}
    \email[E-mail: ]{O.Nachtmann@thphys.uni-heidelberg.de}
\author{F. Nagel}
    \email[E-mail: ]{research@felixnagel.org}

\affiliation{
Institut f\"ur Theoretische Physik, Philosophenweg 16, 69120
Heidelberg, Germany
}


\begin{abstract}
A method is presented for the analysis of the scalar
potential in the general Two-Higgs-Doublet Model.  This allows us to give the
conditions for the stability of the potential and for electroweak symmetry
breaking in this model in a very concise way.  
These results are then applied to two different Higgs potentials
in the literature, namely the MSSM and the Two-Higgs-Doublet potential
proposed by Gunion et al. The known results for these models follow easily as special cases from the
general results.
In particular, for the potential of Gunion et al. we can clarify
the stability and symmetry breaking properties 
of the model with our method.
\end{abstract}


\maketitle
\newpage

\begin{small}
\tableofcontents
\end{small}
\newpage


\section{Introduction}
\label{sec-intr}

The Standard Model~(SM) of particle physics is theoretically consistent and
experimentally successful to date~\cite{Group:2005di, Eidelman:2004wy}.  
The recently observed
neutrino masses are very small and can be neglected in most high-energy
experiments.  Only one ingredient of the~SM, the Higgs boson, has yet to be
discovered.  From direct searches at LEP a lower bound on the Higgs-boson mass
of 114.4~GeV at 95\%~C.L.\ is obtained when the data from the four LEP
collaborations are combined~\cite{Barate:2003sz}. Furthermore, from
measurements of electroweak precision observables at LEP, SLC and NuTeV that
depend on the Higgs-boson mass through radiative corrections and from other
$W$-boson measurements, one obtains (see Tab.~10.2 in~\cite{Group:2005di})
the
prediction \mbox{$m_H = 91^{+45}_{-32}$}~GeV.  Since the one-loop corrections
depend on the Higgs mass only as \mbox{$\log(m_H/m_W)$} the errors are rather
large and the upper error is larger than the lower one.  Such an indirect
determination of a particle mass is very successful in case of the top-quark,
see also Tab.~10.2 of~\cite{Group:2005di}.  However, there the observables have a
quadratic dependence on the mass and are therefore much more predictive.  A
direct discovery of the Higgs boson at the~LHC is presumably possible up to a
mass of about~1~TeV, see e.g.~\cite{Mitsou:2000nv}. One or several Higgs bosons
may be found opening up the direct study of the scalar sector of particle physics.

Despite its experimental success, the~SM is not satisfactory as a fundamental theory
not only because it contains a large number of parameters, that is coupling parameters and
particle masses:
The squared physical, renormalised Higgs-boson mass~$m_H^2$, which we
expect to be of the order of the squared vacuum 
expectation value~$v^2\approx (250\; \text{GeV})^2$ of the
Higgs field, receives large quantum corrections. These corrections 
depend quadratically on the particle masses that the Higgs boson couples to.
This means that the Higgs-boson mass is sensitive to the heaviest particles
of the theory, for instance from physics which alter the high-energy behaviour
at the GUT or the Planck scale.
In principle, the corrections could be much larger
than~$v^2$ but cancel with the squared bare mass so that the difference
gives~\mbox{$m_H^2 \sim v^2$}.  However, such a fine-tuning is usually
considered unnatural. For this so-called {\it naturalness problem} see
e.g.~\cite{Veltman:1980mj}.
A systematic cancellation of quantum corrections to the squared Higgs-boson mass
is provided by supersymmetry~\cite{Wess:1974tw}.
The simplest supersymmetric extension of the SM is the
Minimal Supersymmetric Standard Model~(MSSM)~\cite{Nilles:1983ge}, 
which has been studied extensively in the literature.
In the MSSM one has two Higgs doublets. In further extensions like
the Next-to-Minimal Supersymmetric Standard Model (NMSSM)~\cite{Fayet:1974pd} 
a further Higgs singlet is added.

In this paper we study a class of general models having a scalar sector
with two Higgs doublets. We suppose that the 
\smgroup gauge symmetry holds.
In its simplest
version the fermion content of such a model is assumed to be the same as that
of the~SM.  The same is assumed for the gauge bosons, thereby avoiding to
introduce new fundamental interactions.  In principle, electroweak symmetry breaking~(EWSB) 
works in these
models in a similar way as in the~SM. The Lagrangian contains terms that
consist only of scalar fields without derivatives.  These terms form the
scalar potential at tree-level and are responsible for the stability and symmetry-breaking
pattern of the model.  Further, through their covariant derivatives the scalars
have couplings to the gauge bosons, and through Yukawa interactions they
couple to fermions.  After~EWSB these terms are responsible for the generation of
the gauge-boson and fermion masses, respectively.  With increasing number of
scalar fields the number of parameters in the potential becomes soon very
large.
For instance, as we shall see in Sect.~\ref{sec-thdm}
below, there
are 14~parameters to describe the most general potential with two Higgs
doublets
in contrast to only two parameters for
one doublet.
Therefore the characterisation of the symmetry breaking for
different regions in parameter space becomes increasingly complicated.  We
present a formalism for the analysis of stability and of spontaneous symmetry breaking in models with two Higgs doublets. 

There exists a vast amount of literature on
the~Two-Higgs-Doublet Model~(THDM), where typically the number of parameters of
the potential is restricted by continuous or discrete symmetries.  For
instance in~\cite{Deshpande:1977rw} a detailed discussion of the
symmetry-breaking pattern for different regions in parameter space is given
for the~THDM where a $\mathbbm{Z}_2$~symmetry is imposed on the Higgs
potential and in~\cite{Velhinho:1994vh} the stability of the \CP conserving~THDM 
is investigated.
We also want to mention an approach~\cite{Kastening:1992by} 
to deduce the parameter constraints from stability and 
symmetry breaking in one specific model, the THDM
introduced by Gunion~et~al.~(\cite{Gunion:1989we}, see chapter 4).
Further, in two complementary works~\cite{Ferreira:2004yd,Barroso:2005sm},
the hierarchy between the charge breaking
and the charge conserving minima is investigated for the general THDM.
Basis independent techniques for the general THDM are used in different recent
works~\cite{Branco:2005em,DavidsonHaberONeil:2005,Ivanov:2005hg,Ginzburg:2005yw}
to analyse various aspects of the vacuum.

Here we deduce the parameter constraints from the stability and from the
electroweak symmetry breaking conditions in the general THDM.
The global minimum of the potential is found by
the determination of all stationary points. 
Our results agree with those of~\cite{Deshpande:1977rw} if we impose the
conditions on our parameters such that the potential is invariant under that
discrete symmetry.  Moreover, our formulation of the criteria for stability
and~EWSB of the potential is very concise and should, therefore, be interesting
for its method. This general method, where the potential
is expressed in terms of gauge invariant functions,
was proposed already in a previous work~\cite{Nagel:2004sw}.

We also remark that the scope of the present analysis is the {\em classical}
level.
In a more detailed study quantum corrections should be taken into account.
Some aspects of radiative corrections
for the Higgs potential in constrained $n$-Higgs-Doublet Models are discussed
in~\cite{Flores:pr}. 
The question if stability at the classical level is really
necessary for a consistent quantum theory was put forward a long
time ago by Symanzik~\cite{Symanzik:1973hx}.
The answer given in~\cite{Coleman:1973sx} was that, indeed,
it is necessary. Thus, the results obtained at the classical level
are important for the full theory.

This work is organised as follows: In Sect.~\ref{sec-generalities} we give
general motivations for an extended scalar sector and review
theoretical and experimental constraints on Higgs-boson masses in models with
two Higgs doublets.  In Sect.~\ref{sec-thdm} we present the Lagrangian for
the~THDM.  We introduce our notation for the Higgs potential, which is
expressed in terms of gauge invariant functions of the fields.
In Sect.~\ref{sec-stab} we
analyse the conditions for the stability of the potential.  In
Sect.~\ref{sec-pass} we derive expressions for the location of the stationary
points of the potential.  The conditions derived from spontaneous symmetry breaking
of the electroweak gauge group \mbox{\eweakgroup} down to the
electromagnetic gauge group~\mbox{$U(1)_{\rm em}$} are given in
Sect.~\ref{sec-ssbr}.  In Sect.~\ref{sec-potafter} we specify the potential
after~EWSB in our notation.  Eventually, in Sect.~\ref{sec-exam} the results are
applied to two specific models with two Higgs doublets,
the MSSM and the model of Gunion~et~al~\cite{Gunion:1989we}.
We present our conclusions in Sect.~\ref{sec-conclu}.
In Appendices~\ref{app-orbits} and \ref{app-ndoublets} we discuss the
structure of the space of gauge orbits for the THDM and for the general model
with arbitrary number of Higgs doublets, respectively.


\section{Motivations for an extended \mbox{Higgs sector}}
\label{sec-generalities}

Given the fact that theoretically the mechanism of~EWSB in the SM with one
Higgs doublet is working well and that experimentally not even {\em one}
fundamental scalar particle is discovered yet, what are the motivations to
consider an extended Higgs sector?  
Some reasons are as follows.

A promising candidate for a
theory that solves the naturalness problem and has a higher symmetry than
the~SM is the MSSM~\cite{Nilles:1983ge}, for reviews see
e.g.~\cite{Wess:cp,Martin:1997ns}.
Also the~MSSM contains fundamental Higgs fields that
are responsible for the generation of masses.
As the minimal supersymmetric extension of the SM, the MSSM has
two scalar Higgs doublets, being the minimum for an analytic
superpotential and the absence of triangle anomalies.
An extended Higgs sector can improve gauge-coupling unification at high
scales \cite{Gunion:1998ii}.
In particular, supersymmetric models allow the unification to occur
at a sufficiently high scale consistent with the non-observation
of proton decays~\cite{Dimopoulos:1981yj}.
We remark that supersymmetry imposes many relations
between the parameters of the potential of the most general model with two
doublets.
Cosmology provides an additional reason for a non-minimal Higgs sector. 
The experimental lower bound on the Higgs-boson mass in the~SM,
\mbox{$m_H > 114.4$}~GeV, is too high for the electroweak phase transition in
the early universe to provide the thermal instability that is
necessary for baryogenesis~\cite{Sakharov:dj}, for a review
see~\cite{Bernreuther:2002uj}.
In this respect models with additional scalar particles are
more promising than the~SM, see~\cite{Bernreuther:2002uj}.  
Last but not least, given the large spectrum of
fermion masses and the fact that the fermion-scalar interactions are
responsible for their generation, the idea does not seem too abstruse that
several scalar particles are involved in this mechanism.  There are three
known generations of fermions so why should there exist only one Higgs boson?

A study of the general Higgs sector of a theory possessing the gauge
group~$SU(3)_C \times SU(2)_L \times U(1)_Y$ was presented 
in~\cite{Bernreuther:1998rx}. In the following the strong interaction
gauge group $SU(3)_C$ will play no role, so we shall not mention
it further. In~\cite{Bernreuther:1998rx} the scalar fields, collectively
denoted by~$\chi$, are supposed to transform under 
a general representation of the 
gauge group \mbox{\eweakgroup}.  Such a
representation may be reducible and consists of complex unitary and real
orthogonal parts.  However one can show that without
loss of generality it can be assumed that $\chi$ carries a real orthogonal
representation of \eweakgroup~\cite{Bernreuther:1998rx}.  For the~THDM this
correspondence is demonstrated in Appendix~B of~\cite{Haber:1978jt}.  The
scalar potential is then assumed in~\cite{Bernreuther:1998rx} to have a
non-zero vacuum expectation value
\begin{equation}
\tvec{v} \equiv \langle 0 | \chi | 0\rangle \neq 0
\end{equation}
and to leave the electromagnetic subgroup \mbox{$U(1)_{\rm em}$} unbroken as
usual.  We use the boldface letter here in order to signify that $\tvec{v}$
is, like $\chi$, in general a multi-component vector.  
One can then compute particle masses and couplings for
arbitrary representations of scalars.  However, only some representations are
allowed in order to be in agreement with experimental data.  

One main
restriction originates from the observed high suppression
of flavour-changing neutral currents. 
A way to ensure this in the theory is to require that
all quarks of a given charge receive
their masses from the vacuum expectation value of the same Higgs
boson~\cite{Glashow:1976nt}.  Since we analyse only the scalar potential in
this work and do not specify the Yukawa interactions 
we shall not discuss
this condition further here.  

Very relevant for us here are the consequences for the Higgs sector
obtained
from the accurately measured $\rho$-parameter, which relates the masses of
the $W$ and $Z$~bosons, $m_W$ and $m_Z$, to the weak mixing angle~$\theta_{\rm
  w}$:
\begin{equation}
\label{eq-rhodef}
\rho \equiv \left( \frac{m_W}{\cos \theta_{\rm w} \; m_Z} \right)^2.
\end{equation}
Experimentally the~$\rho$-parameter is very close to one~\cite{unknown:2005em}
and this suggests to require theoretically $\rho=1$ at tree level.
This is indeed the case for the SM. For the most general Higgs model
as studied in~\cite{Bernreuther:1998rx} one finds the following.

It is convenient to extend the real
representation carried by the general Higgs field~$\chi$ 
to a unitary representation of the same (complex) dimension and to
decompose it into representations with definite values~\mbox{$(t,y)$}, 
where $t$ and $y$ are the
weak-isospin and weak-hypercharge quantum numbers, respectively.
We have
\begin{equation}
t = 0,\, \frac{1}{2},\, 1,\, \frac{3}{2},\, \ldots
\end{equation}
and, for reasons discussed in~\cite{Bernreuther:1998rx}, suppose
the hypercharge quantum numbers~$y$ to be rational numbers.
The normalisation is such that the charge, hypercharge and third
component of weak isospin matrices are related by
\begin{equation}
Q=T_3+Y\mbox{.}
\end{equation}
Then the squared
gauge-boson masses, (see~(2.43) in~\cite{Bernreuther:1998rx}), are given by
\begin{align}
\label{eq_mws}
m_W^2 &= \frac{1}{2} \left(\frac{e}{\sin \theta_{\rm w}}\right)^2 \sum_{t,y}
\left[t(t+1) - y^2\right] \tvec{v}^\trans \mathbbm{P}(t,y)
  \tvec{v},\\
\label{eq_mzs}
m_Z^2 &= \left(\frac{e}{\sin \theta_{\rm w} \cos \theta_{\rm w}}\right)^2
\sum_{t,y} y^2\; \tvec{v}^\trans \mathbbm{P}(t,y) \tvec{v},
\end{align}
where \mbox{$\mathbbm{P}(t,y)$} is the projector on the subspace with
representation~\mbox{$(t,y)$}. Here, the positron charge~$e$, and the sine and
cosine of the weak mixing angle are defined in terms of the gauge couplings
$g$ and $g'$ as in the~SM, see for instance~\cite{Nachtmann:1990ta}:
\begin{align}
\sin \theta_{\rm w} &= \frac{g'}{\sqrt{g^2 + g^{\prime \, 2}}},\\
\cos \theta_{\rm w} &= \frac{g}{\sqrt{g^2 + g^{\prime \, 2}}},\\[.2cm]
e &= g \sin \theta_{\rm w},
\end{align}
where we use the same notation as in~\cite{Bernreuther:1998rx}.
It is shown in~\cite{Bernreuther:1998rx} that
\begin{equation}
\tvec{v}^\trans \mathbbm{P}(t,y) \tvec{v} \neq 0
\end{equation}
is only possible if
\begin{equation}
y \in \{-t, -t+1, \ldots, t \}.
\end{equation}
Inserting the expressions for~$m_W$ and~$m_Z$ in the
definition~(\ref{eq-rhodef}) one obtains~\cite{Bernreuther:1998rx}
\begin{equation}
\label{eq-rhogeneral}
\rho = \frac{\sum_{t,y} [t(t+1)-y^2]\; \tvec{v}^\trans
    \mathbbm{P}(t,y) \tvec{v}}{\sum_{t,y} 2y^2\;
    \tvec{v}^\trans \mathbbm{P}(t,y) \tvec{v}}.
\end{equation}
To obtain \mbox{$\rho = 1$} one can either
fine-tune the parameters of the potential in order to get the right vacuum
expectation values, which seems rather unnatural and is therefore
not considered here.  Or one can only allow those representations in
(\ref{eq_mws}) and (\ref{eq_mzs})
that {\em separately} lead to~\mbox{$\rho = 1$}.
There are infinitely many such representations~\cite{Tsao:1980em}, starting
with the doublet with
\mbox{$t = 1/2$} and \mbox{$y = \pm 1/2$}, and the septuplet with \mbox{$t =
3$} and~\mbox{$y = \pm 2$}.  {}From each of these representations one or more
copies are allowed and one still gets \mbox{$\rho = 1$}. Furthermore,
the singlet with $y=0$ and all representations with
\begin{equation}
y \notin \{ -t, -t+1, \ldots, t \}
\end{equation}
can occur because they do not contribute to the sums in~(\ref{eq-rhogeneral}).  

The simplest possibility to extend the Higgs sector of the~SM 
keeping $\rho=1$ at tree level is, therefore, to 
allow for more than one Higgs doublet. 
In these models the shape of the scalar
potential depends on many parameters and can be quite complicated.  As
mentioned above it is the potential that is responsible for the scalar
self-interactions and---together with the interaction terms of the scalars
with the respective particle---for the generation of the masses.  Therefore
one is interested in the conditions that one has to impose on these parameters
in order to render the potential stable and to guarantee spontaneous 
symmetry breaking from \eweakgroup
to~\mbox{$U(1)_{\rm em}$}.  
In this paper we consider two Higgs
doublets, that is the~THDM.
We remark that after EWSB three degrees of freedom in the scalar sector 
reappear as longitudinal modes of the massive
gauge-bosons.  All other degrees of freedom of the scalar sector
correspond to
physical Higgs bosons, that is with each additional doublet four (real)
physical scalar degrees of freedom are added to the model. 
In the THDM, there
are altogether five physical Higgs particles: 
Three neutral Higgs bosons $h^0$, $H^0$
(where conventionally \mbox{$m_{h^0} \leq m_{H^0}$}) and~$A^0$,
as well as two charged Higgs bosons~$H^{\pm}$.
If the Higgs potential is \CP~conserving the neutral
mass eigenstates are also \CP~eigenstates, where 
$h^0$ and~$H^0$ are scalar bosons and~$A^0$ is a pseudoscalar.  
There exist various
studies of the phenomenology of the THDM in the literature,
for an overview and further
references see for instance~\cite{Diaz:2002tp}.  

For the MSSM a large number of
Feynman rules involving Higgs bosons is derived in~\cite{Gunion:1984yn}.  The
phenomenology of the Higgs bosons in the MSSM is further developed
in~\cite{Gunion:1986nh}.  We remark that in models that possess an extended
Higgs sector (and may also contain further non-SM particles) for certain
regions of the parameter space there often exists one neutral Higgs boson that
behaves similarly to the SM~Higgs boson.  For instance, the MSSM
Higgs sector is described by two parameters, which can be chosen as the mass
of the pseudoscalar boson~$m_{A^0}$ and the ratio~\mbox{$\tan \beta$} of the
vacuum expectation values of the two Higgs doublets.
In the decoupling limit \mbox{$m_{A^0} \gg m_Z$},
where practically \mbox{$m_{A^0} \gtrsim 200$}~GeV is
sufficient~\cite{Carena:2002es}, one neutral Higgs boson~$h^0$ is light and has
the same couplings as the SM~Higgs boson whereas the other Higgs bosons $H^0$,
$A^0$ and~$H^{\pm}$ are heavy and decouple.  
If there exist light supersymmetric particles that couple to~$h^0$ it may be comparatively
easy to distinguish $h^0$ from the SM~Higgs boson even if it 
has SM like couplings; this is because $h^0$ can decay into the light
supersymmetric particles if kinematically allowed.
Further, if the light supersymmetric particles couple
to photons (gluons) the one-loop \mbox{$\gamma \gamma h^0$}
(\mbox{$ggh^0$})
coupling is modified by their contribution to the loop,
thus the branching ratios of~$h^0$ differ from those of the SM~Higgs boson.
If all heavy Higgses are
beyond kinematical reach in the decoupling limit, such precision measurements
are the only way to distinguish $h^0$ from the SM~Higgs boson.  Notice that at
an \mbox{$e^+ e^-$}~collider like the ILC~\cite{Aguilar-Saavedra:2001rg}
the heavy Higgs states can only be produced
pairwise in the decoupling limit so that the kinematical limit may be very crucial.  
However, at a \mbox{$\gamma \gamma$}~collider 
s-channel resonant $H^0$ and $A^0$
production~\cite{Asner:2001ia} is possible 
so that only the available c.m. energy of the $\gamma \gamma$~system
limits the masses which can be explored. 
In the $\gamma \gamma$ option of an ILC one expects that the maximal useful
$\gamma \gamma$~c.m. energy will be $80$\% of the c.m.~energy in the
$e^+ e^-$ mode~\cite{Badelek:2001xb}.

Present experiments give the following exclusion regions for 
various versions of the THDM.
The OPAL collaboration has performed a parameter scan for the \CP~conserving
THDM~\cite{Abbiendi:2004gn} and excluded at 95\%~C.L.\ large parts of the
region where
\begin{equation}
\begin{gathered}
\phantom{00}1\text{ GeV} \leq m_{h^0} \leq 130 \text{ GeV},\\
3\text{ GeV} \leq m_{A^0} \leq 2  \text{ TeV},\\
0.4 \leq \tan \beta \leq 40,\\
\alpha =
  -\frac{\pi}{2}, -\frac{\pi}{4}, 0, \frac{\pi}{4}, \frac{\pi}{2}.
\end{gathered}
\end{equation}
Here $\alpha$ is a mixing angle for the two states~$h^0$ and~$H^0$.
Further, the approximate region where
\begin{equation}
\begin{split}
1 \text{ GeV} &< m_{h^0} < 55 \text{ GeV},\\
3 \text{ GeV} &< m_{A^0} < 63 \text{ GeV}
\end{split}
\end{equation}
is excluded 
for all $\tan \beta$ values for negative $\alpha$.
In a combined analysis~\cite{:2001xy} of
the four LEP collaborations a lower bound on the mass of the charged Higgs in
models with two Higgs doublets like the THDM or the MSSM, approximately
\begin{equation}
m_{H^{\pm}} > 78.6 \text{ GeV}
\end{equation}
is determined.  In another analysis~\cite{:2001xx} of the four
LEP collaborations signals for neutral Higgs bosons at different benchmark
points of the~MSSM were searched for.  Here the limits
\begin{equation}
\begin{split}
m_{h^0}  &>  91.0 \text{ GeV},\\
m_{A^0}  &>  91.9 \text{ GeV}
\end{split}
\end{equation}
at 95\%~C.L.\ are obtained.  Under the assumption that 
the ``left-right''-stop mixing
is maximal and with {\em conservative} choices for other MSSM parameters the
region \mbox{$0.5 < \tan \beta < 2.4$} is excluded at~95\%~C.L.


\section{The general Two-Higgs-Doublet Model}
\label{sec-thdm}

We denote the two complex Higgs-doublet fields by
\begin{equation}
\label{eq-doubldef}
\varphi_i(x) = \begin{pmatrix} \varphi^+_i(x) \\  \varphi^0_i(x) \end{pmatrix}
\end{equation}
with \mbox{$i = 1, 2$}.  Hence we have eight real scalar degrees of freedom.
The most general \eweakgroup invariant Lagrangian for
the~THDM can be written as
\begin{equation}
\label{eq-lagr}
\mathscr{L}_{\rm THDM} = \mathscr{L}_{\varphi} + \mathscr{L}_{\rm Yuk} +
\mathscr{L}', 
\end{equation}
where the pure Higgs-boson Lagrangian is given by
\begin{equation}  
\mathscr{L}_{\varphi} = \sum_{i = 1,2}
\left(\mathcal{D}_{\mu} \varphi_i \right)^{\dagger} \left(\mathcal{D}^{\mu}
  \varphi_i \right) - V(\varphi_1, \varphi_2).
\end{equation}
This term replaces the kinetic terms of the Higgs boson and the Higgs
potential in the SM~Lagrangian.  
The covariant derivative is
\begin{equation}
\label{eq-covderiva}
\mathcal{D}_{\mu} = \partial_{\mu} + i g  W^a_{\mu} \mathbf{T}_a + i g' B_{\mu}
\mathbf{Y},
\end{equation}
where $\mathbf{T}_a$ and $\mathbf{Y}$ are the generating operators of
weak-isospin and weak-hypercharge transformations.
For the Higgs doublets we have
 \mbox{$\mathbf{T}_a = \tau_a /
2$}, where $\tau_a$ (\mbox{$a = 1,2,3$})
are the Pauli matrices.
We assume both doublets to have weak hypercharge~\mbox{$y = 1/2$}.
Further, $\mathscr{L}_{\rm Yuk}$ are the Yukawa-interaction terms of the Higgs fields
with fermions.  Finally, $\mathscr{L}'$ contains the terms of the
Lagrangian without Higgs fields.  We do not specify $\mathscr{L}_{\rm Yuk}$
and~$\mathscr{L}'$ here since they are not relevant for our analysis.  The
Higgs potential~$V$ in the~THDM will be specified below and
discussed extensively.

We remark that in the~MSSM the two Higgs doublets $H_1$ and $H_2$ carry
hypercharges \mbox{$y = -1/2$} and \mbox{$y = +1/2$}, respectively, whereas
here we use the conventional definition of the~THDM with both doublets
carrying \mbox{$y = +1/2$}.  However, our analysis can be translated to the
other case, see for example (3.1) in~\cite{Gunion:1984yn}, by setting
\begin{equation}
\label{eq-thdmsusytrafo}
\begin{split}
\varphi^\alpha_{1} & = - \epsilon_{\alpha \beta} ( H_1^{\beta} )^{*},\\
\varphi^\alpha_{2} & = H_2^\alpha,
\end{split}
\end{equation}
where $\epsilon$ is given by
\begin{equation}
\epsilon = \begin{pmatrix} \phantom{-}0 & 1\phantom{-} \\ -1 & 0\phantom{-} \end{pmatrix}.
\end{equation}

The most general gauge invariant and renormalisable potential
  \mbox{$V(\varphi_1,\varphi_2)$} for the two Higgs doublets~$\varphi_1$
  and~$\varphi_2$ is a hermitian linear combination of the following terms:
\begin{equation}
\label{eq-potterms}
\varphi_i^{\dagger}\varphi_j,\quad
 \big( \varphi_i^{\dagger}\varphi_j \big)
 \big(\varphi_k^{\dagger}\varphi_l \big), 
\end{equation}
where \mbox{$i,j,k,l \in \{ 1, 2\}$}.
It is convenient to discuss the properties of the potential such
as its stability and its spontaneous symmetry breaking in terms of
gauge invariant expressions.
For this purpose we arrange the \eweakgroup invariant
scalar products
into the hermitian \mbox{$2 \by 2$}~matrix
\begin{equation}
\label{eq-kmat}
\twomat{K} :=
\begin{pmatrix}
  \varphi_1^{\dagger}\varphi_1 & \varphi_2^{\dagger}\varphi_1 \\
  \varphi_1^{\dagger}\varphi_2 & \varphi_2^{\dagger}\varphi_2
\end{pmatrix}
\end{equation}
and consider its decomposition
\begin{equation}
\label{eq-kmatdecomp}
\twomat{K}_{i j} =
 \frac{1}{2}\,\left( K_0\,\delta_{i j} + K_a\,\sigma^a_{i j}\right),
\end{equation}
using the completeness of the Pauli matrices $\sigma^a$ ($a=1,2,3$), 
together with the unit matrix.
The four real coefficients defined by the decomposition~(\ref{eq-kmatdecomp})
are given by
\begin{equation}
\label{eq-kdef}
K_0 = \varphi_{i}^{\dagger} \varphi_{i},
\qquad
K_a = ( \varphi_{i}^{\dagger} \varphi_{j} )\, \sigma^a_{ij} ,
\quad (a=1,2,3).
\end{equation}
Here and in the following summation over repeated indices is understood.
Using the inversion of~(\ref{eq-kdef}),
\begin{equation}
\label{eq-phik}
\begin{alignedat}{2}
\varphi_1^{\dagger}\varphi_1 &= (K_0 + K_3)/2, &\qquad
\varphi_1^{\dagger}\varphi_2 &= (K_1 + i K_2)/2, \\ 
\varphi_2^{\dagger}\varphi_2 &= (K_0 - K_3)/2, &
\varphi_2^{\dagger}\varphi_1 &= (K_1 - i K_2)/2,
\end{alignedat}
\end{equation}
the most general potential can be written as follows:
\begin{subequations}
\label{eq-vdef}
\begin{align}
V(\varphi_1,\varphi_2) &= V_2 + V_4,\\ 
V_2 &= \xi_0 K_0 + \xi_a K_a, \\
V_4 &= \eta_{00} K_0^2 + 2 K_0 \eta_a K_a + K_a
\eta_{ab} K_b,
\end{align}
\end{subequations}
where the 14 independent parameters $\xi_0$, $\xi_a$, $\eta_{00}$, $\eta_a$
and \mbox{$\eta_{ab}=\eta_{ba}$} are real.
We subsequently write
\mbox{$\tvec{K}:=(K_a)$},
\mbox{$\tvec{\xi}:=(\xi_a)$},
\mbox{$\tvec{\eta}:=(\eta_a)$}
 and
\mbox{$E:=(\eta_{ab})$}.

Now we consider a change of basis of the Higgs fields,
$\varphi_i \rightarrow \varphi'_i$, where
\begin{equation}
\label{eq-udef}
\begin{pmatrix} \varphi'_1 \\
                \varphi'_2 \end{pmatrix}
= \begin{pmatrix} U_{11} & U_{12} \\
                  U_{21} & U_{22} \end{pmatrix}
  \begin{pmatrix} \varphi_1 \\
                  \varphi_2 \end{pmatrix} .
\end{equation}
Here
\begin{equation}
U = \begin{pmatrix} U_{11} & U_{12} \\
                    U_{21} & U_{22} \end{pmatrix},
\qquad
U^\dagger U = \unitmatrix,
\end{equation}
is a \mbox{$2 \by 2$}~unitary transformation.
With~\eqref{eq-udef} the gauge invariant functions~(\ref{eq-kdef}) transform as
\begin{equation}
\label{eq-biltrafo}
K'_0 = K_0,
\qquad
K'_a = R_{ab}(U) K_b,
\end{equation}
where \mbox{$R_{ab}(U)$} is defined by
\begin{equation}
U^\dagger \sigma^a U = R_{ab}(U)\,\sigma^b.
\end{equation}
The matrix~\mbox{$R(U)$} has the properties
\begin{equation}
\label{eq-rprodet}
R^\ast(U)=R(U),
\quad
R^\trans(U)\, R(U) = \unitmatrix,
\quad
\det R(U) = 1,
\end{equation}
where $\unitmatrix$ denotes the \mbox{$3 \by 3$}~unit matrix.
That is, $R(U)\in SO(3)$.
The form of the Higgs potential~(\ref{eq-vdef}) remains unchanged under
the replacement~(\ref{eq-biltrafo}) if we perform an appropriate transformation
of the parameters
\begin{equation}
\label{eq-partrafo}
\begin{alignedat}{2}
\xi'_0 &= \xi_0,  & \tvec{\xi}' &= R(U)\,\tvec{\xi}, \\
\eta'_{00} &= \eta_{00}, &  \tvec{\eta}' &= R(U)\,\tvec{\eta}, \\
 E' &= R(U)\,E\,R^\trans(U).
\end{alignedat}
\end{equation}
Moreover, for every matrix~$R$ with the properties~(\ref{eq-rprodet}),
there is a unitary transformation~(\ref{eq-udef}).
We can therefore diagonalise~$E$, thereby reducing the number of parameters
of~$V$ by three.  The Higgs potential is then determined by only 11 real
parameters.

The matrix~$\twomat{K}$ is positive semi-definite, which
follows immediately from its definition~(\ref{eq-kmat}).
With \mbox{$K_0 = \trace{\twomat{K}}$} and
\mbox{$K_0^2 - \tvec{K}^2 = 4 \det{\twomat{K}}$} this implies
\begin{equation}
\label{eq-kinq}
K_0 \geq 0,\qquad K_0^2 - \tvec{K}^2 \geq 0.
\end{equation}
On the other hand, for any given $K_0, \tvec{K}$
fulfilling~(\ref{eq-kinq}),
it is possible to find fields $\varphi_i$ obeying~(\ref{eq-kdef}).
Furthermore, all fields obeying~\eqref{eq-kdef} for a given
$K_0, \tvec{K}$ form one gauge orbit.
This is shown explicitly in Appendix~\ref{app-orbits}.

Thus, the functions~$K_0, K_a$ parametrise the gauge orbits
and not a unique Higgs-field configuration.
Specifying the domain of the functions~$K_0, K_a$ corresponding
to the gauge orbits allows to discuss the potential
directly in the form~(\ref{eq-vdef})
with all gauge degrees of freedom eliminated.
It is curious to note that
the gauge orbits of the Higgs fields
of the THDM are parametrised by Minkowski type four vectors $(K_0,\tvec{K})$
which have to lie on or inside the forward light cone.

In the following sections we derive bounds on the parameters of the potential
that result from the conditions that
\begin{itemize}
\item the potential~$V$ is stable,
\item we have spontaneous symmetry breaking of \eweakgroup
  down to \emgroup .
\end{itemize}


\section{Stability}
\label{sec-stab}

According to Sect.~\ref{sec-thdm} we can analyse the properties of the
potential~(\ref{eq-vdef}) as a function of~$K_0$ and~$\tvec{K}$ on the domain
determined by \mbox{$K_0 \geq 0$} and \mbox{$K_0^2 \geq \tvec{K}^2$}.  For
\mbox{$K_0 > 0$} we define
\begin{equation}
\label{eq-ksde}
\tvec{k} := \tvec{K} / K_0.
\end{equation}
In fact, we have \mbox{$K_0 = 0$} only for
\mbox{$\varphi_1 = \varphi_2 = 0$},
and the potential is $V=0$ in this case.
From~(\ref{eq-vdef}) and~(\ref{eq-ksde}) we obtain for~\mbox{$K_0 > 0$}
\begin{align}
\label{eq-vk}
V_2 &= K_0\, J_2(\tvec{k}),&
J_2(\tvec{k}) &:= \xi_0 + \tvec{\xi}^\trans \tvec{k},\\
\label{eq-vk4}
V_4 &= K_0^2\, J_4(\tvec{k}),&
J_4(\tvec{k}) &:= \eta_{00} 
  + 2 \tvec{\eta}^\trans \tvec{k} + \tvec{k}^\trans E \tvec{k},
\end{align}
where we introduce the functions $J_2(\tvec{k})$ and $J_4(\tvec{k})$
on the domain \mbox{$|\tvec{k}| \leq 1$}.

For the potential~$V$ to be stable, it must be bounded from below. The stability is determined by the behaviour of $V$
in the limit $K_0 \to \infty$, governed by $J_4(\tvec{k})$ and $J_2(\tvec{k})$.

\begin{enumerate}[leftmargin=0pt, itemindent=*, labelwidth=\textwidth, align=left]

\item[(a)] If there exists $\tvec{k}$ with $|\tvec{k}|\le1$ such that
\[
J_4(\tvec{k})<0,
\]
the potential is unstable.

\item[(b)] Now consider the case
\[
J_4(\tvec{k})\ge0
\qquad \text{for all } \tvec{k}\text{ with }|\tvec{k}|\le1.
\]

\begin{enumerate}[leftmargin=0pt, itemindent=*, labelwidth=\textwidth, align=left]

\item[(b.1)] If there exists $\tvec{k}$ with $|\tvec{k}|\le1$ such that
\[
J_4(\tvec{k})=0 \; \text{ and } \; J_2(\tvec{k})<0,
\]
the potential is unstable.

\item[(b.2)] If for all~$\tvec{k}$ with $|\tvec{k}|\le1$ we have
\begin{equation} \label{eq-jinq}
J_4(\tvec{k})>0,
\end{equation}
the potential is stable. We call this case \emph{stability in the strong sense}. To prove the stability for this case we write the potential $V$~\eqref{eq-vdef} using~\eqref{eq-vk} and \eqref{eq-vk4} as follows
\[
V(K_0, \tvec{k}) = K_0 J_2(\tvec{k}) + K_0^2 J_4(\tvec{k}).
\]
The functions $J_2(\tvec{k})$ and $J_4(\tvec{k})$ are continuous on the compact domain $|\tvec{k}|\le1$. Therefore, they attain their maximum and minimum values there:
\[
J_{2\, \mathrm{min}} \le J_2(\tvec{k}) \le J_{2\, \mathrm{max}},\quad J_{4\, \mathrm{min}} \le J_4(\tvec{k}) \le J_{4\, \mathrm{max}}.
\]
Note that in the case~(b.2) we have $J_{4\, \mathrm{min}} > 0$. Hence we obtain, since $K_0 \ge 0$,
\begin{eqnarray}
\lefteqn{\!\!\!\!\!\!\!V(K_0, \tvec{k}) \;\ge\; K_0 J_{2\, \mathrm{min}} + K_0^2 J_{4\, \mathrm{min}}}\nonumber\\
 & = & J_{4\, \mathrm{min}} \left( K_0 + \frac{J_{2\, \mathrm{min}}}{2 J_{4\, \mathrm{min}}} \right)^2 - \frac{1}{4} \frac{(J_{2\, \mathrm{min}})^2}{J_{4\, \mathrm{min}}}\label{eq-pot-b2}\\
 & \ge & - \frac{1}{4} \frac{(J_{2\, \mathrm{min}})^2}{J_{4\, \mathrm{min}}}.\nonumber
\end{eqnarray}
This proves that $V$ is bounded from below in the case~(b.2), and thus $V$ is stable.

Now consider $V$ on an arbitrary path in~$K$ space going to infinity
\[
(K_0, \tvec{k}) = (K_0, \tvec{k}(K_0)),\quad |\tvec{k}(K_0)|\le1,\quad K_0 \to \infty.
\]
The second line in~\eqref{eq-pot-b2} yields
\begin{equation}
\label{eq-limit}
V(K_0, \tvec{k}(K_0)) \to \infty \qquad \text{for } K_0 \to \infty   .
\end{equation}

\item[(b.3)] If for all~$\tvec{k}$ with $|\tvec{k}|\le1$ we have
\begin{equation} \label{eq-j4j2stab}
J_4(\tvec{k})>0,
\qquad \text{or}\qquad
J_4(\tvec{k})=0 \text{ and } J_2(\tvec{k})>0,
\end{equation}
with $J_4(\tvec{k})=0$ occurring for some $\tvec{k}$,
the potential is stable. We call this case \emph{stability in the weak sense}.

To prove that $V$ is stable we define the compact region $B_0$ as
\[
B_0 = \{ \tvec{k} \;;\; |\tvec{k}|\le1,\; J_4(\tvec{k})=0 \}.
\]
According to~(b.3) we have $J_2(\tvec{k})>0$ for $\tvec{k} \in B_0$. Since $J_2(\tvec{k})$ is continuous, it has a positive minimum on~$B_0$:
\[
J_2(\tvec{k}) \ge J_{2\, \mathrm{min}, B_0} > 0.
\]
Further we define the compact domains
\[
B_1 = \big\{ \tvec{k} \;;\; |\tvec{k}| \le 1,\; J_2(\tvec{k})\ge \tfrac{1}{2} J_{2\, \mathrm{min}, B_0} \big\}
\]
and, if $B_1$ is not equal to the complete region $|\tvec{k}|\le1$,
\[
B_2 = \big\{ \tvec{k} \;;\; |\tvec{k}| \le 1,\; J_2(\tvec{k}) \le \tfrac{1}{2} J_{2\, \mathrm{min}, B_0} \big\}.
\]
For $\tvec{k} \in B_1$ we have
\[
J_4(\tvec{k}) \ge 0 \; \text{ and } \; J_2(\tvec{k}) \ge \tfrac{1}{2} J_{2\, \mathrm{min}, B_0}   .
\]
Thus we get here
\begin{equation}
\label{eq-pot-b3-2}
V(K_0, \tvec{k}) \ge K_0 \; \tfrac{1}{2} J_{2\, \mathrm{min}, B_0}.
\end{equation}
Note that it may happen that $B_2$ is empty, for example in the case $J_4(\tvec{k}) \equiv 0$ where
\[
B_0 = B_1 = \{ \tvec{k} \;;\; |\tvec{k}|\le1 \}.
\]

If $B_2$ is non-empty, then $B_0$ and $B_2$ have no points $\tvec{k}$ in common. Then we have for $\tvec{k} \in B_2$
\[
J_4(\tvec{k}) > 0.
\]
Since $B_2$ is compact, $J_4(\tvec{k})$ has a positive minimum on~$B_2$. Therefore the same argument as in case~(b.2), restricted to $B_2$, shows that $V$ is bounded from below on~$B_2$. This concludes the proof that $V$ is bounded from below if the weak-stability conditions are satisfied.

Moreover we find that in the case (b.3) for all paths in $K$ space going to infinity again \eqref{eq-limit} holds, similarly to the situation in~(b.2). To prove this we just have to consider the minimum of the lower bounds on $V(K_0, \tvec{k})$ as function of~$K_0$ given above for the two subcases of~(b.3), that is \eqref{eq-pot-b3-2}, and \eqref{eq-pot-b2} applied to $\tvec{k} \in B_2$ instead of $|\tvec{k}|\le1$.

\item[(b.4)] If for all~$\tvec{k}$ with $|\tvec{k}|\le1$ we have
\[
J_4(\tvec{k})>0,
\qquad \text{or}\qquad
J_4(\tvec{k})=0 \text{ and } J_2(\tvec{k})\ge0 ,
\]
where $J_4(\tvec{k})=0$ together with $J_2(\tvec{k})=0$ occurs for some $\tvec{k}$,
we speak of the \emph{marginal case}.
For potentials in this marginal case, stability holds if and only if there exists a constant~$C>0$ such that
\begin{equation}
\label{eq-pot-b4}
J_2(\tvec{k})^2 \le C\,J_4(\tvec{k})
\end{equation}
for all~$\tvec{k}$ of the region~$B_3$ defined as
\[
B_3 = \{ \tvec{k} \;;\; |\tvec{k}|\le1,\; J_4(\tvec{k})>0,\; J_2(\tvec{k})<0 \}.
\]
We call this \emph{marginal stability}\footnote{We thank J. Tooby-Smith for drawing attention to a missing condition for the stability in the marginal case~\cite{Tooby-Smith:2026kzj}, which we have now included.}.
To prove the foregoing statements we note that for fixed $\tvec{k} \in B_3$ the potential $V(K_0, \tvec{k})$ has, as function of $K_0 \ge 0$, a minimum at
\[
K_{0, \mathrm{min}}(\tvec{k}) = -\frac{1}{2} \frac{J_2(\tvec{k})}{J_4(\tvec{k})},
\]
with value 
\begin{align}
V_{\mathrm{min}}(\tvec{k}) = -\frac{1}{4} \frac{(J_2(\tvec{k}))^2}{J_4(\tvec{k})}.
\end{align}
Hence $V_{\mathrm{min}}(\tvec{k})$ is bounded from below if and only if \eqref{eq-pot-b4} is satisfied for all $\tvec{k} \in B_3$. This proves that, in the marginal case, \eqref{eq-pot-b4} is necessary and sufficient for marginal stability.\\

We have thus separated the different cases (a) and (b.1) to (b.4) of the THDM potential as needed in the following in our paper. As already mentioned, the error in the original version affects only the marginal case.\\

The conditions for stability can be formulated in other, equivalent, ways. One compact form is the following: The potential is stable if and only if for all $\tvec{k}$ with $|\tvec{k}|\le1$ we have either
\[
J_4(\tvec{k})\ge0 \; \text{ and } \; J_2(\tvec{k})\ge0,
\]
or
\[
J_2(\tvec{k})^2 \le C\,J_4(\tvec{k})
\]
with a constant $C > 0$ independent of~$\tvec{k}$.\\ 

Finally we emphasize that in the discussion of the cases (a) and (b) we have also shown the following: The potential satisfies~\eqref{eq-limit} for all paths in $K$ space going to infinity if and only if $V$ is stable in the strong or weak sense.
\end{enumerate}
\end{enumerate}

To assure $J_4(\tvec{k})$ is positive (semi-)definite, it is sufficient to
consider its value for all stationary points of \mbox{$J_4(\tvec{k})$}
on the domain \mbox{$\abs{\tvec{k}} < 1$}, and for all stationary points on the
boundary~\mbox{$|\tvec{k}| = 1$}. This holds, because
the global minimum of the continuous function \mbox{$J_4(\tvec{k})$}
is reached on the compact domain $\abs{\tvec{k}}\le 1$, and it is among those
stationary points.
This leads to bounds on $\eta_{00}$, $\eta_{a}$ and $\eta_{ab}$, which
parametrise the quartic term~$V_4$ of the potential.  For $\abs{\tvec{k}}< 1$
the stationary points---if there are any---must fulfil
\begin{equation}
\label{eq-cklo}
E \tvec{k} = - \tvec{\eta} \quad \text{with }\abs{\tvec{k}} < 1 .
\end{equation}
If \mbox{$\det E \neq 0$} we explicitly obtain
\begin{equation}
\left. J_4(\tvec{k}) \right|_{\substack{stat}} =
  \eta_{00} - \tvec{\eta}^\trans E^{-1} \tvec{\eta}
 \quad \text{if } \;1 - \tvec{\eta}^\trans E^{-2} \tvec{\eta} > 0,
\end{equation}
where the inequality follows from the condition \mbox{$\abs{\tvec{k}} < 1$}.
If \mbox{$\det E = 0$} there can exist one or more ``exceptional'' solutions
$\tvec{k}$ of~(\ref{eq-cklo}).
They, again, have to obey \mbox{$\abs{\tvec{k}}<1$}.
For \mbox{$\abs{\tvec{k}} = 1$} we must find the stationary points of the
function
\begin{equation}
\label{eq-fdef}
F_4(\tvec{k},u) := J_4(\tvec{k}) + u \left(1 - \tvec{k}^2\right),
\end{equation}
where $u$ is a Lagrange multiplier.  Those are given by
\begin{equation}
\label{eq-soll}
(E - u) \tvec{k} = -\tvec{\eta} \quad \text{with }\abs{\tvec{k}} = 1.
\end{equation}
For regular values of~$u$ such that \mbox{$\det (E - u) \neq 0$}
the stationary points are given by
\begin{equation}
\label{eq-kaol}
\tvec{k}(u) = -(E - u)^{-1} \tvec{\eta}, 
\end{equation}
and the Lagrange multiplier is determined from the condition
\mbox{$\tvec{k}^\trans \tvec{k} = 1$} after inserting~(\ref{eq-kaol}):
\begin{equation}
\label{eq-lmul}
1 - \tvec{\eta}^\trans (E - u)^{-2} \tvec{\eta} = 0.
\end{equation}
We thus obtain the solution
\begin{equation}
\left. J_4(\tvec{k}) \right|_{\substack{stat}} = u + \eta_{00} - \tvec{\eta}^\trans
(E - u)^{-1} \tvec{\eta},
\end{equation}
where $u$ is a solution of~(\ref{eq-lmul}).  Also for \mbox{$\abs{\tvec{k}}=1$},
depending on the parameters $\eta_a$ and $\eta_{ab}$, there can be
exceptional solutions \mbox{$(\tvec{k}, u)$} of~(\ref{eq-soll}) where
\mbox{$\det(E - u) = 0$}, i.e.\ where $u$ is an eigenvalue of $E$.

The regular solutions for the two cases \mbox{$\abs{\tvec{k}} < 1$} and
\mbox{$\abs{\tvec{k}} = 1$} can be described using one function only.
Considering~(\ref{eq-fdef}) and~(\ref{eq-kaol}), we define
\begin{equation}
f(u) := F_4(\tvec{k}(u), u),
\end{equation}
with \mbox{$\tvec{k}(u)$} as in~(\ref{eq-kaol}).  This leads to
\begin{align}
\label{eq-flam}
f(u) & =  u + \eta_{00} - \tvec{\eta}^\trans (E - u)^{-1} \tvec{\eta},\\
\label{eq-flampr}
f'(u) & =  1 - \tvec{\eta}^\trans (E - u)^{-2} \tvec{\eta},
\end{align}
so that for all ``regular'' stationary points~$\tvec{k}$
of~$J_4(\tvec{k})$
\begin{align}
f(u) &= \left. J_4(\tvec{k}) \right|_{\substack{stat}},\\
f'(u) &= 1 - \tvec{k}^2
\end{align}
holds, where we set $u=0$ for the solution with $\abs{\tvec{k}}<1$.
There are stationary points of~$J_4(\tvec{k})$ with $\abs{\tvec{k}}<1$
and $\abs{\tvec{k}}=1$ exactly if $f'(0)>0$ and $f'(u)=0$, respectively,
and the value of $J_4(\tvec{k})$ is then given by $f(u)$.

\begin{figure}
\includegraphics[width=\linewidth]{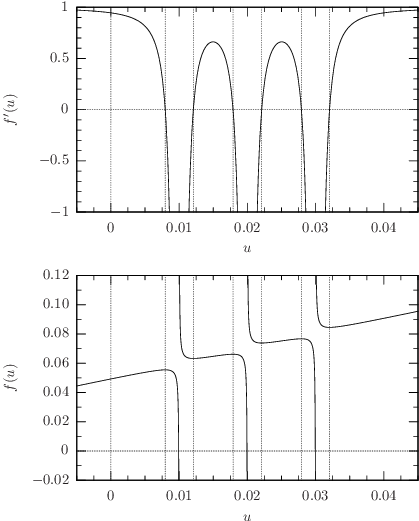}
\caption{\label{fig:functions}
The stability determining functions $f'(u)$ and $f(u)$
as given by~(\protect{\ref{eq-fprd}}) and~(\protect{\ref{eq-fdiag}})
with \mbox{$\eta_{00} = 0.05$},
     \mbox{$(\mu_1,\mu_2, \mu_3) = (0.01, 0.02, 0.03)$} and
     \mbox{$(\eta_1, \eta_2, \eta_3) = (0.002, 0.002, 0.002)$}.
}

\end{figure}

In a basis where \mbox{$E = {\rm diag}(\mu_1, \mu_2, \mu_3)$} we obtain:
\begin{align}
\label{eq-fdiag}
f(u) &= u + \eta_{00} - \sum_{a = 1}^3 \frac{\eta_a^2}{\mu_a - u},\\
\label{eq-fprd}
f'(u) &= 1 - \sum_{a = 1}^3 \frac{\eta_a^2}{(\mu_a - u)^2}.
\end{align}
The derivative~\mbox{$f'(u)$} has at most 6 zeros.
The shape of~\mbox{$f(u)$} and~\mbox{$f'(u)$} for a (purely didactical) set of parameters
where $f'(u)$ has 6 zeros can be seen in Fig.~\ref{fig:functions}.
Notice that there are no exceptional solutions
if in this basis all three components of~$\tvec{\eta}$ are different from zero.

The function $f(u)$ given by~\eqref{eq-flam} allows us to discuss also the
exceptional solutions of~\eqref{eq-cklo} and \eqref{eq-soll}.
Consider first $\abs{\tvec{k}}<1$ and suppose that $\det E=0$.
In the basis where $E$ is diagonal we have
\begin{equation}
\det E = \mu_1\, \mu_2\, \mu_3 = 0
\end{equation}
and~\eqref{eq-cklo} reads
\begin{equation}
\label{eq-stabesol}
\begin{split}
\mu_1\, k_1 &= -\eta_1,\\
\mu_2\, k_2 &= -\eta_2,\\
\mu_3\, k_3 &= -\eta_3.
\end{split}
\end{equation}
Clearly, a solution of~\eqref{eq-stabesol} is only possible if with
$\mu_a=0$ also $\eta_a=0$ ($a=1,2,3$).
Therefore, we see from~\eqref{eq-fdiag} that exceptional solutions with
$\abs{\tvec{k}}<1$ are only possible if $f(u)$ stays finite at $u=0$.
That is, the pole which would correspond to $\mu_a=0$ must have residue zero.
Suppose now that indeed $\eta_a=0$ for all~$a$ where $\mu_a=0$.
Take as an example $\mu_1=\mu_2=0$ and $\eta_1=\eta_2=0$, but $\mu_3 \neq 0$.
Then we get the general solution of~\eqref{eq-stabesol} as
\begin{equation}
k_3 = -\frac{\eta_3}{\mu_3},
\end{equation}
with $k_1$, $k_2$ arbitrary but satisfying
\begin{equation}
\tvec{k}^2 = k_1^2+k_2^2+\left(\frac{\eta_3}{\mu_3}\right)^2 < 1 .
\end{equation}
We can write this as
\begin{equation}
\label{eq-stabessol}
\tvec{k} = \tvec{k}_\parallel + \tvec{k}_\perp,
\end{equation}
where
\begin{gather}
\begin{split}
\label{eq-stabescon}
\tvec{k}_\parallel = -\frac{1}{\mu_3}\tvec{\eta},
\qquad
E\, \tvec{k}_\perp = 0,\\
\tvec{k}_\perp^2 < 1 - \tvec{k}_\parallel^2 = 1-\left(\frac{\eta_3}{\mu_3}\right)^2.
\end{split}
\end{gather}
For the functions~\eqref{eq-fdiag}, \eqref{eq-fprd} we get here
\begin{align}
f(u) &= u + \eta_{00} - \frac{\eta_3^2}{\mu_3 - u},\\
f'(u) &= 1 - \frac{\eta_3^2}{(\mu_3 - u)^2}.
\end{align}
Inserting the solution~$\tvec{k}$ from~\eqref{eq-stabessol},
\eqref{eq-stabescon} in $J_4(\tvec{k})$
we get
\begin{align}
\label{eq-j4exinner}
f(0) &= \left. J_4(\tvec{k}) \right|_{\substack{stat}},\\
f'(0) &= 1 - \tvec{k}_\parallel^2 > \tvec{k}_\perp^2 \geq 0 .
\end{align}
Clearly these arguments work similarly, if only one of the $\mu_a$ is equal
to zero or all three $\mu_a$ are zero.
In all cases~\eqref{eq-j4exinner} holds for the exceptional points with
$\abs{\tvec{k}}<1$, which can exist only if $f(u)$ has no pole at $u=0$.
Since~\eqref{eq-j4exinner} involves only ''scalar'' quantities,
it holds in any basis.

The case of exceptional solutions for $\abs{\tvec{k}}=1$ can be treated in
an analogous way. An exceptional solution of~\eqref{eq-soll} with $u=\mu_a$
($a=1,2,3$) can only exist if the corresponding $\eta_a=0$.
Then $f(u)$ has no pole for $u=\mu_a$ and the exceptional solutions
of~\eqref{eq-soll} fulfil
\begin{equation}
\tvec{k} = \tvec{k}_\parallel + \tvec{k}_\perp,
\end{equation}
with
\begin{equation*}
\tvec{k}_\parallel = -\left.(E-u)^{-1} \tvec{\eta} \right|_{\substack{u=\mu_a}},\quad
(E - \mu_a) \tvec{k}_\perp = 0
\end{equation*}
and
\begin{align}
f(\mu_a) &= \left. J_4(\tvec{k}) \right|_{\substack{stat}},\\
f'(\mu_a) &= 1 - \tvec{k}_\parallel^2 = \tvec{k}_\perp^2 \ge 0.
\end{align}
Note that if a solution is possible, $\tvec{k}_\perp$ may be any linear
combination of the eigenvectors to the eigenvalue~$\mu_a$ of $E$, where the norm
is given by~$\abs{\tvec{k}_\perp} = \sqrt{f'(\mu_a)}$.
Thus we see that the function $f(u)$
is very useful for discussing the stability of the THDM potential.
What we have shown so far can be formulated as follows.

Consider the functions~$f(u)$ and $f'(u)$.
Denote by $I$,
\begin{equation}
\label{eq-idef}
I = \{ u_1, \dots, u_n \},
\end{equation}
the following set of values of $u$.
Include in $I$ all $u$ where $f'(u)=0$. Add $u=0$ to $I$ if $f'(0)>0$.
Consider then the eigenvalues $\mu_a$ ($a=1,2,3$) of $E$.
Add those $\mu_a$ to $I$ where $f(\mu_a)$ is finite and $f'(\mu_a) \geq 0$.
We have $n \leq 10$. The values of the function
$J_4(\tvec{k})$
at its stationary points are given by
\begin{equation}
\left. J_4(\tvec{k}) \right|_{\substack{stat}} = f(u_i)
\end{equation}
with $u_i \in I$. The potential is stable if  $f(u_i)>0$ for all $u_i \in I$.
Then the stability is given solely by the quartic terms in the potential.
The potential is unstable if we have $f(u_i)<0$ for at least one $u_i \in I$.
If we have $f(u_i)\geq 0$ for all $u_i \in I$ and $f(u_i)=0$ for at least one
$u_i \in I$ we have to consider in addition
$J_2(\tvec{k})$
in order to decide on the stability of the potential.

We turn now to this latter case. We shall show that we have then to consider
in addition the function
\begin{equation}
\label{eq-gdef}
g(u) := \xi_0 - \tvec{\xi}^\trans (E-u)^{-1} \tvec{\eta}.
\end{equation}
for the stationary points of~$J_4(\tvec{k})$ with
\begin{equation}
\label{eq-v4flat}
\left.J_4(\tvec{k})\right|_{\substack{stat}} = f(u_i) =0.
\end{equation}
We have for the vectors~$\tvec{k}$ satisfying~\eqref{eq-v4flat}
\begin{equation}
\label{eq-j2grineq}
J_2(\tvec{k}) = g(u_i)
\end{equation}
if $u_i \neq \mu_a$, that is $u_i$ is not an eigenvalue of $E$.
If $u_i$ is an eigenvalue of $E$, that is $u_i = \mu_a \in I$,
and $\tvec{e}_l(u_i)$ ($l=1,\ldots,N$) are the $N\le 3$~eigenvectors to~$u_i$,
then we have
\begin{equation}
\label{eq-j2geineq}
\inf_{\tvec{k}} J_2(\tvec{k}) = g(u_i) - \abs{\tvec{\xi}_\perp (u_i)} \sqrt{f'(u_i)},
\end{equation}
where the infimum is taken over all exceptional solutions~$\tvec{k}$ to~$u_i$
and
\begin{equation}
\label{eq-xip}
\tvec{\xi}_\perp (u_i) :=
  \sum_{l=1}^{N} \frac{\tvec{\xi}\tvec{e}_l(u_i)}{\abs{\tvec{e}_l(u_i)}^2}\, \tvec{e}_l(u_i).
\end{equation}
We summarise our findings in a theorem.

\begin{theorem}
\begin{samepage}
\label{theorem-stability}
The most general potential of the two-Higgs-doublet model has the
form~\eqref{eq-vdef}.
Its stability is decided in the following way.
If the potential has only the quadratic term $V_2$,
it is stable for $\xi_0 > \abs{\tvec{\xi}}$,
marginally stable for $\xi_0 = \abs{\tvec{\xi}}$
and unstable for $\xi_0 < \abs{\tvec{\xi}}$.
Suppose now that $V_4 \not\equiv 0$.
We construct then the functions $f(u)$~\eqref{eq-flam}, $f'(u)$~\eqref{eq-flampr}
and $g(u)$~\eqref{eq-gdef},
and the set $I$~\eqref{eq-idef} of (at most 10) $u$ values.
\begin{enumerate}
\item
If $f(u_i)>0$ for all $u_i \in I$ the potential is stable in the strong
sense~\eqref{eq-jinq}.
\item
If $f(u_i)<0$ for at least one $u_i \in I$ the potential is unstable.
\item
If $f(u_i) \geq 0$ for all $u_i \in I$ and $f(u_i)=0$ for at least one
$u_i \in I$ we consider also the function $g(u)$~\eqref{eq-gdef}.
The potential is stable in the weak sense~\eqref{eq-j4j2stab}
if for all $u_i \in I$ where $f(u_i)=0$ the
following holds (see~\eqref{eq-j2grineq} to \eqref{eq-xip}):
\begin{alignat}{2}
\label{eq-gstabregular}
g(u_i)&>0 &\;
  &\text{if}\; u_i \neq \mu_a,\\
\label{eq-gstabexcept}
\quad g(u_i)-\abs{\tvec{\xi}_\perp (u_i)}\sqrt{f'(u_i)}&>0 &
  &\text{if}\; u_i = \mu_a.
\end{alignat}
If for some or all $u_i \in I$ where $f(u_i)=0$ we have in~\eqref{eq-gstabregular}  and/or \eqref{eq-gstabexcept} $=0$ instead of $>0$ we have the marginal case. If at least for one $u_i \in I$ where $f(u_i)=0$ we have in~\eqref{eq-gstabregular}  and/or \eqref{eq-gstabexcept} $<0$ instead of $>0$ the potential is unstable. The marginal case is not decided by this criterion alone and has to be checked using the marginal-stability condition stated above.
\end{enumerate}
\end{samepage}
\end{theorem}

Our theorem gives a complete characterisation of the stability properties
of the general THDM potential.
In the following subsection we apply the theorem to
assert the strong stability condition~\eqref{eq-jinq} for a specific potential.
An application for the weaker stability condition~\eqref{eq-j4j2stab}
is given in Subsection~\ref{ssec-mssmex}.


\subsection{Stability for THDM of Gunion et al.}

We consider the~THDM of~\cite{Gunion:1989we} with the Higgs potential
\begin{equation}
\begin{split}
\label{eq-gunp}
&V(\varphi_1,\varphi_2) = \lambda_1 \big(\varphi_1^{\dagger} \varphi_1 -
v_1^2\big)^2 + \lambda_2 \big(\varphi_2^{\dagger} \varphi_2 - v_2^2\big)^2\\ 
&\quad + \lambda_3 \left(\varphi_1^{\dagger} \varphi_1 -
v_1^2 + \varphi_2^{\dagger} \varphi_2 - v_2^2\right)^2\\
&\quad + \lambda_4
\left(\big(\varphi_1^{\dagger} \varphi_1\big)\big(\varphi_2^{\dagger}
  \varphi_2\big) -
  \big(\varphi_1^{\dagger}\varphi_2\big)\big(\varphi_2^{\dagger}\varphi_1\big)
\right)\\
&\quad + \lambda_5 \left({\rm Re} \big(\varphi_1^{\dagger}\varphi_2\big)
  - v_1 v_2 \cos \xi\right)^2\\
&\quad + \lambda_6 \left({\rm Im}
\big(\varphi_1^{\dagger}\varphi_2\big) - v_1 v_2 \sin \xi\right)^2\\ 
&\quad + \lambda_7 \left({\rm Re} \big(\varphi_1^{\dagger}\varphi_2\big) -
  v_1 v_2 \cos \xi\right) \left({\rm Im}
  \big(\varphi_1^{\dagger}\varphi_2\big) - v_1 v_2 \sin \xi\right),
\end{split}
\end{equation}
which contains nine real parameters if we do not count the constant.
This potential breaks the discrete symmetry
\begin{equation}
\label{eq-dissym}
\varphi_1 \longrightarrow -\varphi_1, \qquad
\varphi_2 \longrightarrow \varphi_2
\end{equation}
only softly, i.e.~by $V_2$ terms,
thus suppressing large flavour changing neutral currents.
For various restrictions on the THDM by
symmetries see for instance~\cite{Diaz:2002tp}.  Dropping the constant term,
we put the potential into the form~(\ref{eq-vdef}) using the
relations~(\ref{eq-phik}).  Then,
\begin{equation}
\begin{split}
\label{eq-pagu}
\eta_{00}
  &= \frac{1}{4}(\lambda_1 + \lambda_2 + 4 \lambda_3 + \lambda_4),\\[.2cm]
\tvec{\eta}
  &= \frac{1}{4} \begin{pmatrix} 0\\ 0\\ \lambda_1 - \lambda_2 \end{pmatrix},
 \\[.2cm]
E &= \frac{1}{8} \begin{pmatrix}
    2(\lambda_5-\lambda_4) & \lambda_7 & 0\\
    \lambda_7 & 2(\lambda_6-\lambda_4) & 0\\
    0 & 0 & 2(\lambda_1+\lambda_2-\lambda_4) \end{pmatrix}.
\end{split}
\end{equation}
From~(\ref{eq-flam}) and~(\ref{eq-flampr}) we obtain
\begin{align}
\begin{split}
f(u) &= u + \frac{1}{4} (\lambda_1 + \lambda_2 + 4
\lambda_3 + \lambda_4)\\
&\quad- \frac{\left(\lambda_1 - \lambda_2 \right)^2}
  { 4(\lambda_1 + \lambda_2 - \lambda_4 - 4 u) },\\
f'(u) &= 1 - \frac{\left(\lambda_1 - \lambda_2 \right)^2}
  {\left( \lambda_1 + \lambda_2 - \lambda_4 - 4 u \right)^2}. 
\end{split}
\end{align}
We introduce the abbreviation
\begin{equation}
\label{eq-abkapp}
\kappa := \frac{1}{2} \left(\lambda_5 + \lambda_6 -
\sqrt{(\lambda_5-\lambda_6)^2 +\lambda_7^2}\right).
\end{equation}
Applying theorem~\ref{theorem-stability}
to the functions \mbox{$f(u)$} and~\mbox{$f'(u)$},
we find the strong stability assertion by $V_4$, see~\eqref{eq-jinq},
to be equivalent to the simultaneous conditions
\begin{equation}
\begin{gathered}
\label{eq-gunstab}
\lambda_1 + \lambda_3 > 0,
\qquad
\lambda_2 + \lambda_3 > 0,\\
\lambda_4, \kappa >
-2\lambda_3 - 2\sqrt{(\lambda_1 + \lambda_3)(\lambda_2 + \lambda_3)} .
\end{gathered}
\end{equation}
In particular, if \mbox{$\lambda_1, \lambda_2, \lambda_3, \lambda_4,
\kappa > 0$} these inequalities are fulfilled.
They can then be rewritten as:
\begin{equation}
\label{eq-partcond}
\lambda_1, \lambda_2, \lambda_3, \lambda_4 > 0, \qquad 4\lambda_5
\lambda_6 > \lambda_7^2.
\end{equation}
That means, if~(\ref{eq-partcond}) holds, the conditions~(\ref{eq-gunstab})
are fulfilled and the potential is stable.  For the case
\mbox{$\lambda_7 = 0$} we can replace $\kappa$
 by~$\min(\lambda_5,\lambda_6)$ in the stability
conditions~(\ref{eq-gunstab}), which are then
in particular fulfilled if \mbox{$\lambda_i > 0$} for
\mbox{$i = 1, \ldots, 6$}.

The potential in~\cite{Deshpande:1977rw} is even more specific since it is
invariant under~(\ref{eq-dissym}).
Inserting their potential parameters in~\eqref{eq-gunstab}
we reproduce the result of~\cite{Deshpande:1977rw}, their equation~(2).


\section{Location of stationary points}
\label{sec-pass}
In the following we only consider potentials which fulfill the strong or the weak stability condition. 
After our stability analysis in the preceding section we now determine the
location of the stationary points of the potential, since among these points
there are the local and global minima.  To this end we define
\begin{equation}
\fvec{K} = \begin{pmatrix} K_0\\ \tvec{K} \end{pmatrix},\quad
\fvec{\xi} = \begin{pmatrix} \xi_0\\ \tvec{\xi} \end{pmatrix},\quad
\fmat{E} = \begin{pmatrix} \eta_{00} & \tvec{\eta}^\trans\\
                           \tvec{\eta} & E \end{pmatrix}.
\end{equation}
In this notation the potential~(\ref{eq-vdef}) reads
\begin{equation}
\label{eq-vtil}
V = \fvec{K}^\trans \fvec{\xi} + \fvec{K}^\trans
\fmat{E} \fvec{K}
\end{equation}
and is defined on the domain
\begin{equation}
\label{eq-domv}
\fvec{K}^\trans \fmat{g} \fvec{K} \geq 0,
\qquad K_0 \ge 0,
\end{equation}
with
\begin{equation}
\fmat{g} = \begin{pmatrix} 1 & \phantom{-}0 \\ 0 & -\unitmatrix \end{pmatrix}.
\end{equation}
For the discussion of the stationary points of~$V$,
we distinguish the allowed cases
$\fvec{K}=0$,
$K_0 > \abs{\tvec{K}}$
and
$K_0 = \abs{\tvec{K}} > 0$.

The trivial configuration $\fvec{K}=0$
is a stationary point of the potential with~$V=0$,
as a direct consequence of the definitions.

The stationary points of~$V$ in the inner part of the domain,
$K_0>\abs{\tvec{K}}$,
are given by
\begin{equation}
\label{eq-statin}
\fmat{E} \fvec{K} = - \frac{1}{2} \fvec{\xi}
\quad
\text{with}
\quad
\fvec{K}^\trans \fmat{g} \fvec{K}>0
\quad
\text{and}
\quad
K_0>0.
\end{equation}
For \mbox{$\det \fmat{E} \neq 0$} we obtain the unique solution
\begin{equation}
\label{eq-unisol}
\fvec{K} = - \frac{1}{2} \fmat{E}^{-1} \fvec{\xi},
\end{equation}
provided that
\begin{equation}
\label{eq-incon}
\fvec{\xi}^{\, \rm T} \fmat{E}^{-1} \fmat{g} \fmat{E}^{-1}
\fvec{\xi} > 0
\quad
\text{and}
\quad
K_0
> 0,
\end{equation}
and no solution if (\ref{eq-incon}) does not hold.
The Hessian matrix
\begin{equation}
\label{eq-secder}
\left( \frac{\partial^{\,\! 2}}{\partial K_i \, \partial K_j}\, V \right)
= 2 \fmat{E},
 \quad \text{where} \quad i,j = 0 \dots 3,
\end{equation}
determines whether (\ref{eq-unisol}) is a local minimum, a local maximum or a
saddle.  In the case \mbox{$\det \fmat{E} = 0$} we may have exceptional
solutions of~(\ref{eq-statin}).
In the regular case as well as in the exceptional cases,
the existence of a solution of~\eqref{eq-statin}
along with the corresponding values of the potential are not affected by the
transformation of parameters~\eqref{eq-partrafo}.

The stationary points of $V$ on the domain boundary
$K_0 = \abs{\tvec{K}} > 0$ are
stationary points of the function
\begin{equation}
\tilde{F}\big(\fvec{K}, u \big) := V - u \fvec{K}^\trans
\fmat{g} \fvec{K},
\end{equation}
where $u$ is a Lagrange multiplier.  The relevant stationary points of
\mbox{$\tilde{F}$} are given by
\begin{equation}
\label{eq-stap}
\big(\fmat{E} - u \fmat{g} \big) \fvec{K} =
-\frac{1}{2} \fvec{\xi}
\quad \text{with }\fvec{K}^\trans \fmat{g} \fvec{K}=0 \text{ and } K_0 >0.
\end{equation}
For regular values of $u$ with
 \mbox{${\rm det} (\fmat{E}-u \fmat{g} ) \neq 0$}
we obtain
\begin{equation}
\label{eq-solktil}
\fvec{K}(u) =
 -\frac{1}{2} \big(\fmat{E}-u \fmat{g}\big)^{-1} \fvec{\xi}.
\end{equation}
The Lagrange multiplier is determined from
the constraints in~(\ref{eq-stap}) by inserting~\eqref{eq-solktil}:
\begin{equation}
\label{eq-lagm}
\fvec{\xi}^{\, \rm T} \big(\fmat{E}-u \fmat{g}\big)^{-1}
\fmat{g} \big(\fmat{E}-u \fmat{g}\big)^{-1} \fvec{\xi} = 0
\quad \text{and} \quad K_0 >0.
\end{equation}
There may be up to 4 values \mbox{$u = \tilde{\mu}_a$} with \mbox{$a =
1,\ldots,4$} for which \mbox{${\rm det} (\fmat{E}-u \fmat{g}) = 0$}.
Depending on the potential some or all of them may lead to exceptional
solutions of~(\ref{eq-stap}).
Note that for the regular as well as for the exceptional cases,
the Lagrange multipliers~$u$ and the value of the potential
belonging to solutions
\mbox{$(u, \fvec{K})$} of~(\ref{eq-stap})
are, similar to above, invariant under the transformations~\eqref{eq-partrafo}.

For \emph{any} stationary point the potential is given by
\begin{equation}
\label{eq-statexpl}
V|_{\substack{stat}} = \frac{1}{2} \fvec{K}^\trans \fvec{\xi} = -
\fvec{K}^\trans \fmat{E} \fvec{K}.
\end{equation}
Suppose now that the weak stability condition~(\ref{eq-j4j2stab})
holds.
Then~\eqref{eq-statexpl} gives for non-trivial stationary points
where~\mbox{$\fvec{K} \neq 0$}:
\begin{equation}
\label{eq-statneg}
V|_{\substack{stat}}
  < 0,
\end{equation}
since the cases $V_4 < 0$ and $V_4=V_2=0$ are excluded by the stability condition.

Similarly to the stability analysis in Sect.~\ref{sec-stab} we can
use a unified description for the regular stationary points of~$V$ with
$K_0>0$ for both \mbox{$|\tvec{K}| < K_0$} and \mbox{$|\tvec{K}| = K_0$}
defining the function
\begin{equation}
\tilde{f}(u) := \tilde{F}
\big(\fvec{K}(u), u\big),
\end{equation}
where \mbox{$\fvec{K}(u)$} is the solution~(\ref{eq-solktil}).
It follows:
\begin{align}
\label{eq-ftil}
\tilde{f}(u) &= -\frac{1}{4} \fvec{\xi}^{\, \rm T}
 \big( \fmat{E} - u \fmat{g} \big)^{-1} \fvec{\xi},\\
\label{eq-ftilpr}
\tilde{f}'(u) &= - \frac{1}{4} \fvec{\xi}^{\, \rm T}
\big( \fmat{E}-u \fmat{g} \big)^{-1} \fmat{g}
  \big( \fmat{E}- u \fmat{g} \big)^{-1} \fvec{\xi}.
\end{align}
Denoting the first component of \mbox{$\fvec{K}(u)$} as $K_0(u)$
we summarise as follows.

\begin{theorem}
\begin{samepage}
\label{classes-statpoints}
The stationary points of the potential are given by
\begin{itemize}
\item[(I\,a)]
$\fvec{K} = \fvec{K}(0)$
if $\tilde{f}'(0) < 0$,\  $K_0(0)>0$ and \mbox{$\det \fmat{E} \neq 0$},
\item[(I\,b)]
solutions $\fvec{K}$ of~\eqref{eq-statin}
if \mbox{$\det \fmat{E} = 0$},
\item[(II\,a)]
$\fvec{K}=\fvec{K}(u)$
for $u$ with \mbox{$\det (\fmat{E} - u \fmat{g}) \neq 0$},\ 
 \mbox{$\tilde{f}'(u) = 0$} and
 $K_0(u)>0$,
\item[(II\,b)]
solutions $\fvec{K}$ of~\eqref{eq-stap}
for $u$ with \mbox{$\det (\fmat{E} - u\fmat{g}) = 0$},
\item[(III)]
$\fvec{K}$ = 0.
\end{itemize}
\end{samepage}
\end{theorem}

In many cases, for instance if all values
\mbox{$\tilde{\mu}_1, \ldots, \tilde{\mu}_4$} are different,
we can diagonalise the in general non-hermitian matrix
\mbox{$\fmat{g} \fmat{E}$} in the following way:
\begin{equation}
\fmat{g} \fmat{E} = \sum_{a=1}^{4} \tilde{\mu}_a \tilde{\mathbbm{P}}_a.
\end{equation}
Here the $\tilde{\mathbbm{P}}_a$ are quasi-projectors constructed from
the normalised right and left eigenvectors $\chi_a,\tilde{\chi}_a$
of $\fmat{g} \fmat{E}$. We have then
$
\fmat{g} \fmat{E}\, \chi_a = \tilde{\mu}_a\, \chi_a,\;
\tilde{\chi}_a\, \fmat{g} \fmat{E} = \tilde{\chi}_a \, \tilde{\mu}_a,\;
\tilde{\chi}_a \chi_b = \delta_{a b}
$
and can impose as additional normalisation condition
\mbox{$\chi_a^\dagger \chi_a = 1$}.
The $\tilde{\mathbbm{P}}_a$ are given by
\begin{equation}
\tilde{\mathbbm{P}}_a = \chi_a \tilde{\chi}_a
\end{equation}
and satisfy
\begin{equation}
\trace \tilde{\mathbbm{P}}_a = 1, \qquad
\tilde{\mathbbm{P}}_a \tilde{\mathbbm{P}}_b =
  \begin{cases}  \tilde{\mathbbm{P}}_a & \text{for $a=b$},\\
                 0                     & \text{for $a \neq b$}, \end{cases}
\end{equation}
where \mbox{$a, b \in \{ 1, \ldots, 4\}$}. In terms of the $\tilde{\mathbbm{P}}_a$
(\ref{eq-ftil}) and~(\ref{eq-ftilpr}) read
\begin{align}
\tilde{f}(u) &= - \frac{1}{4} \sum_{a=1}^4
  \frac{\fvec{\xi}^\trans \tilde{\mathbbm{P}}_a \, \fmat{g} \, \fvec{\xi}}
       {\tilde{\mu}_a - u},\\ 
\tilde{f}'(u) &= - \frac{1}{4} \sum_{a=1}^4
  \frac{\fvec{\xi}^\trans \tilde{\mathbbm{P}}_a \, \fmat{g} \, \fvec{\xi}}
       {(\tilde{\mu}_a - u)^2}.
\end{align}
Of course, $\tilde{f}(u)$~\eqref{eq-ftil} is always a meromorphic
function of $u$ but in general poles of higher order than one may also occur.


\section{Criteria for electroweak symmetry breaking}
\label{sec-ssbr}

The global minimum will be among the stationary points discussed in
the previous section.
Here we discuss the spontaneous symmetry breaking features of the
possible classes of minima and give criteria to ensure a global minimum
with the required electroweak symmetry breaking
$\eweakgroup \rightarrow \emgroup$.

A global minimum at $\fvec{K}=0$ means vanishing fields for the vacuum.
In this case, no symmetry is spontaneously broken.
If the global minimum lies at $\fvec{K}\neq0$, the full gauge group or
a subgroup is broken.
We denote the vacuum expectation values,
i.e.~the fields at the global minimum of the potential~$V$, by
\begin{equation}
v_i^+ := \langle \varphi_i^+ \rangle,
\qquad
v_i^0 := \langle \varphi_i^0 \rangle,
\end{equation}
with $i=1,2$. In general the $v_i^+, v_i^0$ are complex numbers.
To exhibit the consequences of electromagnetic gauge invariance we consider
the matrix~(\ref{eq-kmat}) at the global minimum,
$\twomat{K}|_{\rm min}$.

If the global minimum of~$V$ occurs with \mbox{$K_0 > |\tvec{K}|$},
it follows \mbox{$\det \twomat{K} |_{\rm min} > 0$},
see Sect.~\ref{sec-thdm}. Since we have
\begin{equation}
\det \twomat{K} |_{\rm min} =
    \left| v_1^+\; v_2^0 - v_1^0\; v_2^+ \right|^2,
\end{equation}
the vectors
\begin{equation}
\label{eq-vvec}
\begin{pmatrix} v_1^+\\ v_2^+\end{pmatrix},
\qquad
\begin{pmatrix} v_1^0\\ v_2^0\end{pmatrix}
\end{equation}
are linearly independent.  Then there is no transformation~(\ref{eq-udef})
such that both $v_1^+$ and~$v_2^+$ become zero.  This means that the
full gauge group \eweakgroup is broken.
We can also show this using the methods explained in Appendix~\ref{app-orbits},
see~\eqref{eq-phivacfull}.

In the case that the global minimum of~$V$ features \mbox{$K_0 = |\tvec{K}|>0$},
the rank of the matrix \mbox{$\twomat{K} |_{\rm min}$} is 1 and the
vectors~(\ref{eq-vvec}) are linearly dependent. After performing a
\eweakgroup transformation we achieve
\begin{align}
\label{eq-aglo}
\begin{pmatrix} v_1^+\\ v_2^+\end{pmatrix} &= 0,\\
\label{eq-gaugeneutral}
\begin{pmatrix} v_1^0\\ v_2^0\end{pmatrix} &=
\begin{pmatrix} \abs{v_1^0}\phantom{\,e^{i\zeta}} \\  
                \abs{v_2^0}\,e^{i\zeta}  \end{pmatrix}
\neq 0,
\qquad \zeta \in \realnums,
\end{align}
and identify the unbroken~\mbox{$U(1)$} gauge group with the electromagnetic
one.  By a transformation~(\ref{eq-udef}), namely
\begin{align}
\label{eq-goldrot}
\begin{pmatrix} \varphi_1' \\ \varphi_2' \end{pmatrix}
=
\begin{pmatrix}  \cos\beta               & \sin\beta\, e^{-i \zeta } \\
                -\sin\beta\, e^{i \zeta} & \cos\beta  \end{pmatrix}
\begin{pmatrix} \varphi_1 \\ \varphi_2 \end{pmatrix},
\end{align}
with $\beta$ fulfilling $\abs{v_1^0} \sin\beta = \abs{v_2^0} \cos\beta$,
we can arrange that 
\begin{align}
\label{eq-aglop}
\begin{pmatrix} v_1'^+\\ v_2'^+\end{pmatrix} &= 0,\\
\label{eq-atra}
\begin{pmatrix} v_1'^0 \\        v_2'^0  \end{pmatrix} &= 
\begin{pmatrix} v_0/\sqrt{2} \\ 0 \end{pmatrix},
\qquad v_0 > 0
\end{align}
for the vacuum expectation values in the new basis.
This can also be derived from the results in Appendix~\ref{app-orbits},
see~\eqref{eq-kmatr1evals} et seq.
In~\eqref{eq-atra} $v_0$ is the usual Higgs-field vacuum expectation
value, $v_0 \approx 246$~GeV (see for instance~\cite{Nachtmann:1990ta}).

Now, we want to derive conditions for the parameters in the
general potential~(\ref{eq-vdef}), which lead to
the required EWSB by a global minimum with $K_0=|\tvec{K}|>0$.
In the following, we assume the potential to be stable.
If we consider parameters fulfilling \mbox{$\xi_0 \geq |\tvec{\xi}|$}
this immediately implies 
\mbox{$J_2(\tvec{k})\ge 0$} and hence from~\eqref{eq-j4j2stab} 
$V>0$ for all \mbox{$\fvec{K} \neq 0$}.
Therefore for these parameters the global minimum is at~$\fvec{K}=0$.
Thus we arrive at the requirement
\begin{equation}
\label{eq-cond}
\xi_0 < |\tvec{\xi}|.
\end{equation}
Here we obtain
\begin{equation}
\left. \frac{\partial V}{\partial K_0} \right|_{
  \begin{subarray}{l} \tvec{k}\;\text{fixed},\\
                    K_0 = 0 \end{subarray}
} = \xi_0 +  \tvec{\xi}^\trans \tvec{k}
  < 0 
\end{equation} 
for some $\tvec{k}$, i.e.\ the global minimum of $V$ lies
at~\mbox{$\fvec{K} \neq 0$}.

Addressing the non-trivial cases,
suppose that the two points
\begin{equation}
\fvec{p} = \begin{pmatrix} p_0 \\ \tvec{p} \end{pmatrix},\qquad
\fvec{q} = \begin{pmatrix} q_0 \\ \tvec{q} \end{pmatrix}
\end{equation}
with \mbox{$p_0 \geq \abs{\tvec{p}}$} and \mbox{$q_0 \geq \abs{\tvec{q}}$}
are stationary points of~$V$, that is each of them is either
a solution of~(\ref{eq-statin}), or, together
with an appropriate Lagrange multiplier $u_p$ or $u_q$
for $\fvec{p}$ or $\fvec{q}$, respectively,
a solution of~(\ref{eq-stap}).

Firstly, consider $p_0 = \abs{\tvec{p}}$.
From~\eqref{eq-vtil} and \eqref{eq-stap} we have

\begin{equation}
\left. \frac{\partial V}{\partial K_0} \right|_{
  \begin{subarray}{l} \tvec{K}\;\text{fixed},\\
                      \fvec{K}=\fvec{p} \end{subarray}
}
 = \xi_0 + 2 (\fmat{E}\,\fvec{p})_0
 = 2 u_p\, p_0.
\end{equation}
If $u_p<0$, there are points $\fvec{K}$ with $K_0>p_0$,
$\tvec{K}=\tvec{p}$ and lower potential in the neighbourhood of
$\fvec{p}$, which therefore can not be a minimum.
We conclude that in a theory with the required EWSB the global
minimum (which needs to be on the light cone) must have
Lagrange multiplier $u_0 \geq 0$.
As we shall show in Sect.~\ref{sec-potafter}, the case $u_0=0$
leads to zero mass for the physical charged Higgs field
at tree level. This is unacceptable from a phenomenological point
of view if we disregard the possibility of very large radiative
corrections.
Therefore, we find as condition for an acceptable theory
\begin{equation}
\label{eq-upositive}
u_0 > 0
\end{equation}
for the Lagrange multiplier $u_0$ of the global minimum.
Secondly, for $p_0 = \abs{\tvec{p}}$ and $q_0 = \abs{\tvec{q}}$ we have
from~\eqref{eq-statexpl} and \eqref{eq-stap}:
\begin{align}
V(\fvec{p})-V(\fvec{q})
&= \frac{1}{2} \fvec{p}^\trans \fvec{\xi} - \frac{1}{2} \fvec{q}^\trans \fvec{\xi}\nonumber\\
&= \fvec{p}^\trans (u_q\, \fmat{g} - \fmat{E} ) \fvec{q}
- \fvec{q}^\trans (u_p\, \fmat{g} - \fmat{E} ) \fvec{p}\nonumber\\
\label{eq-vdiffstaton}
&=(u_q-u_p)\,\fvec{p}^\trans \fmat{g} \fvec{q}.
\end{align}
Since $\fvec{p}$ and $\fvec{q}$ are vectors on the forward light cone,
 $\fvec{p}^\trans \fmat{g} \fvec{q}$ is always non-negative and
zero only for $\fvec{p}$ parallel to $\fvec{q}$.
Furthermore, the case that two different $\fvec{p},\fvec{q}$ are parallel
can not occur, since then \eqref{eq-vdiffstaton} requires $V(\fvec{p})=V(\fvec{q})$,
while \eqref{eq-statexpl} and \eqref{eq-statneg} imply
$V(\fvec{p}) \neq V(\fvec{q})$ for that case.
Therefore we conclude
\begin{equation}
u_p > u_q \qquad \Longleftrightarrow \qquad V(\fvec{p}) < V(\fvec{q}).
\end{equation}

Assuming $p_0 = \abs{\tvec{p}}$ and $q_0 > \abs{\tvec{q}}$ we
get from~\eqref{eq-statexpl} and \eqref{eq-statin}:
\begin{equation}
\label{eq-twostat}
V(\fvec{p})-V(\fvec{q}) = -u_p\,\fvec{p}^\trans \fmat{g} \fvec{q}
\end{equation}
and
\begin{equation}
V(\fvec{p})-V(\fvec{q}) = (\fvec{p}-\fvec{q})^\trans\,\fmat{E}\,(\fvec{p}-\fvec{q}).
\end{equation}
The first equation implies in particular, that a stationary point
on the domain boundary
with positive Lagrange multiplier
will have a lower potential than
any stationary point with $K_0 > \tvec{K}$.
From the second equation follows in this case that $\fmat{E}$ has
a negative eigenvalue.
Since for the stationary point~$\fvec{q}$ in the interior of the light cone
the Hessian matrix is $2\fmat{E}$ (see~\eqref{eq-secder}), we see that
$\fvec{q}$ cannot be a local minimum.
This result and the hierarchies of the stationary points derived above
agree with~\cite{Barroso:2005sm}.
We summarise as follows.

\begin{theorem}
\begin{samepage}
\label{criteria-globalminima}
A global minimum with the spontaneous electroweak symmetry breaking
\mbox{\eweakgroup $\rightarrow$ \emgroup}
and absence of zero mass physical charged \mbox{Higgses}
\begin{itemize}
\item[(I)] requires $\xi_0 < \abs{\tvec{\xi}}$,
\item[(II)] is given and guaranteed by the stationary point of the
classes~(IIa) or (IIb) of theorem~\ref{classes-statpoints} with the
largest Lagrange multiplier $u_0>0$.
\end{itemize}
\end{samepage}
\end{theorem}

We remark that for two different stationary points in the inner part of
the domain or with $u=0$ on its boundary,
any linear combination of them with $K_0 \geq \abs{\tvec{K}}$
is a stationary point as well.
These points therefore belong to one connected set
of degenerate stationary points.
Stability requires that this set contains points with $K_0>\abs{\tvec{K}}$
and is bounded by points with $K_0=\abs{\tvec{K}}$.
If interpreted geometrically, this degenerate set is a line segment,
ellipsoidal area or volume.
Together with the arguments above we find the following
\emph{mutually exclusive} possibilities for local minima,
expressed in terms of the gauge invariant functions:
one or multiple solutions with the required EWSB ($K_0=\abs{\tvec{K}}$)
\emph{or}
the aforementioned degenerate set of solutions
($K_0 \geq \abs{\tvec{K}}$)
\emph{or}
one charge breaking solution ($K_0 > \abs{\tvec{K}}$)
\emph{or}
the trivial solution ($\fvec{K}=0$)
.

\section{Potential after electroweak symmetry breaking}
\label{sec-potafter}

We assume a stable potential which leads to the desired symmetry breaking
pattern as discussed in the previous sections and derive consequences
for the resulting physical fields in the following.
We choose a unitary gauge and the basis for the scalar fields such that for
the vacuum expectation values relations~\eqref{eq-aglop} and~\eqref{eq-atra}
hold, and furthermore the fields satisfy
\begin{align}
\varphi_1^+(x) &= 0,\\
\mIm \varphi_1^0(x) &= 0,\\
\mRe \varphi_1^0(x) &\geq 0.  
\end{align}
We introduce as usual a shifted Higgs field
\begin{equation} 
\rho'(x) := \sqrt{2}\, \mRe \varphi_1^0(x) - v_0.  
\end{equation}
Then the two Higgs doublets are
\begin{equation} 
\varphi_1(x) = \frac{1}{\sqrt{2}}
  \begin{pmatrix} 0\\ v_0 + \rho'(x) \end{pmatrix},
\quad
\varphi_2(x) =
  \begin{pmatrix} \varphi_2^+(x)\\ \varphi_2^0(x) \end{pmatrix}.  
\end{equation}
In addition to~$\rho'$ there are two more neutral Higgs fields
\begin{equation} 
h' := \sqrt{2}\, \mRe \varphi_2^0,
\qquad
h^{\prime \prime} := \sqrt{2}\, \mIm \varphi_2^0 
\end{equation}
and the charged fields
\begin{equation} 
H^+ := \varphi_2^+ ,
\qquad
H^- := \left( H^+ \right)^{\ast}.  
\end{equation}
It is convenient to decompose $\fvec{K}$ according to the power of
the \emph{physical} fields they contain
\begin{equation}
\label{eq-k012}
\fvec{K} = \fvec{K}_\oind{0} + \fvec{K}_\oind{1} +
\fvec{K}_\oind{2}, 
\end{equation}
with
\begin{gather}
\label{eq-kdec1}
\fvec{K}_\oind{0}
  = \begin{pmatrix} v_0^2/2 \\ 0 \\ 0 \\ v_0^2/2\end{pmatrix},\qquad
\fvec{K}_\oind{1}
  = v_0 \begin{pmatrix} \rho'\\ h_0' \\ h_0'' \\ \rho' \end{pmatrix},\\[.4cm]
\label{eq-kdec2}
\fvec{K}_\oind{2} = \frac{1}{2}\begin{pmatrix}
    \rho^{\prime\,2} + 2 H_- H_+ + h^{\prime\, 2} + h^{\prime \prime\, 2} \\
    2 \rho' h' \\
    2 \rho' h'' \\
    \rho^{\prime\,2} - 2 H_- H_+ - h^{\prime\, 2} - h^{\prime \prime\, 2}
  \end{pmatrix}.
\end{gather}
By~$u_0$ we denote again the Lagrange multiplier corresponding to the
global minimum of~$V$.  {}From~(\ref{eq-stap}) we have
\begin{equation}
\label{eq-kabm}
\fmat{E} \fvec{K}_\oind{0} = u_0\, \fmat{g} \fvec{K}_\oind{0} -
\frac{1}{2} \fvec{\xi}.  
\end{equation}
From the explicit expressions~(\ref{eq-kdec1}) and~(\ref{eq-kdec2}) we further
have
\begin{equation}
\label{eq-kexp}
\fvec{K}_\oind{0}^\trans\, \fmat{g}\, \fvec{K}^{\phantom{\trans}}_\oind{0} = 0,
\quad
\fvec{K}_\oind{0}^\trans\, \fmat{g}\, \fvec{K}^{\phantom{\trans}}_\oind{1} = 0.  
\end{equation}
Using~(\ref{eq-k012}) to~(\ref{eq-kexp}) we obtain for the
potential~(\ref{eq-vtil})
\begin{equation} 
V = V_\oind{0} + V_\oind{2} + V_\oind{3} + V_\oind{4}, 
\end{equation}
where $V_\oind{k}$ are the terms of $k^\text{\scriptsize th}$ order in the physical Higgs fields
\begin{align}
\label{eq-v0ex}
V_\oind{0} &= (\xi_0 + \xi_3)\, v_0^2/4,\\
\label{eq-secord}
V_\oind{2}
  &= \fvec{K}_\oind{1}^\trans\, \fmat{E}\, \fvec{K}^{\phantom{\trans}}_\oind{1}
+ 2\, u_0\, \fvec{K}_\oind{0}^\trans\, \fmat{g}\, \fvec{K}^{\phantom{\trans}}_\oind{2},\\
V_\oind{3}
  &= 2\, \fvec{K}_\oind{1}^\trans\, \fmat{E}\, \fvec{K}^{\phantom{\trans}}_\oind{2},\\
V_\oind{4}
  &= \fvec{K}_\oind{2}^\trans\, \fmat{E}\, \fvec{K}^{\phantom{\trans}}_\oind{2}.  
\end{align}
The second order terms~(\ref{eq-secord}) determine the masses of the physical
Higgs fields:
\begin{equation}
\label{eq-mass}
V_\oind{2}
  =\frac{1}{2} (\rho', h', h^{\prime \prime}) {\cal M}_{\rm neutral}^2
               \begin{pmatrix} \rho'\\h'\\h^{\prime \prime} \end{pmatrix}
   + m_{H^\pm}^2 H^+H^-
\end{equation}
with
\begin{align}
\label{eq-mneu}
{\cal M}_\text{neutral}^2 &= 2 \begin{pmatrix}
  -\xi_0 - \xi_3 & -\xi_1 & -\xi_2\\
  -\xi_1 & v_0^2\,(u_0 + \eta_{11}) & v_0^2\, \eta_{12} \\
  -\xi_2 & v_0^2\, \eta_{12} & v_0^2\,(u_0 + \eta_{22}) \end{pmatrix},\\[2ex]
\label{eq-mcha}
m_{H^\pm}^2 &= 2\, u_0\, v_0^2.
\end{align}
Note that the condition $u_0>0$ corresponds to the positivity of
the charged Higgs mass squared at tree level.
This result was already mentioned in Section~\ref{sec-ssbr}.
Generically the mass terms~(\ref{eq-mass}) contain 7 real parameters.
{}From~(\ref{eq-mneu}) and~(\ref{eq-mcha}) we see that all 7 parameters are in general independent in this model.


\section{Examples}
\label{sec-exam}

Here we apply the general considerations of Sections~\ref{sec-stab}
to~\ref{sec-potafter} to specific models.


\subsection{MSSM Higgs potential}
\label{ssec-mssmex}

In this subsection, we consider the MSSM Higgs potential
and reproduce the well-known results
for its stability, symmetry breaking and mass spectrum
(see e.g.~\cite{Martin:1997ns} and references therein),
employing the method described in the previous sections.
In the notation of~\cite{Skands:2003cj} the MSSM Higgs potential is
\begin{equation}
V   = V_D + V_F + V_{\text{soft}}
\end{equation}
with
\begin{equation}
\begin{split}
V_D &=
 \frac{1}{8}(g^2+g'^{\,2}) \left( H_1^\dag H_1 -H_2^\dag H_2 \right)^2
+ \frac{1}{2}g^2 \abs{ H_1^\dag H_2 }^2, \\
V_F &= \abs{\mu}^2 ( H_1^\dagger H_1 +H_2^\dagger H_2), \\
V_{\text{soft}} &= m_{H_1}^2\, H_1^\dagger H_1 + m_{H_2}^2\, H_2^\dagger H_2
- \left( m_3^2\, H_1^\trans \epsilon H_2  + h.c. \right) ,
\end{split}
\end{equation}
where $H_1$ and $H_2$ are Higgs doublets with weak hypercharges
$y=-1/2$ and $y=+1/2$, respectively,
$m_{H_1}^2$, $m_{H_2}^2 \in \realnums$,
$m_3^2 \in \complexnums$ and $\abs{\mu}^2\in\realnums^+_0$
are parameters of dimension mass squared.
Substituting $H_1$ and $H_2$ by doublets $\varphi_1, \varphi_2$ with the
same weak hypercharge $y=+1/2$ according to~(\ref{eq-thdmsusytrafo})
and using the relations~(\ref{eq-phik}), we can put the potential in the
form~(\ref{eq-vdef}).
The parameters are
\begin{equation}
\label{eq-mssmp4}
\eta_{00} = \frac{1}{8}g^2,
  \quad
  \tvec{\eta} = \begin{pmatrix}0\\ 0\\ 0 \end{pmatrix},
  \quad
  E = \frac{1}{8}\begin{pmatrix}-g^2&0&0\\ 0&-g^2&0\\ 0&0&g'^{\,2} \end{pmatrix}
\end{equation}
for $V_4=V_D$ and
\begin{equation}
\label{eq-mssmp2}
\xi_0 = \abs{\mu}^2 + \frac{1}{2}( m_{H_1}^2 + m_{H_2}^2 ),
\quad
\tvec{\xi} =
 \begin{pmatrix} -\mRe{(m_3^2)},\\ \phantom{-}\mIm{(m_3^2)},\\
                 \frac{1}{2}( m_{H_1}^2 - m_{H_2}^2 ) \end{pmatrix}
\end{equation}
for $V_2=V_F+V_{soft}$.

We determine the stability of the potential by employing
theorem~\ref{theorem-stability}.
The functions $f(u)$~\eqref{eq-flam} and $f'(u)$~\eqref{eq-flampr}
for the MSSM are
\begin{align}
f(u) &= u + \frac{1}{8} g^2 ,\\
f'(u) &= 1 .
\end{align}
The set $I$~\eqref{eq-idef} is given here by $u=0$ and the
eigenvalues of $E$~\eqref{eq-mssmp4},
\begin{equation}
I = \left\{ \, u_1 = 0,\;
           u_2 = -\frac{1}{8} g^2,\;
           u_3 =  \frac{1}{8} g'^{\,2} \, \right\} .
\end{equation}
We find for the stationary points of $J_4$
with $u_i=u_1,u_3$ the values $J_4(\tvec{k})|_{\substack{stat}} = f(u_i) > 0$
but for those with $u_2$ the value $J_4(\tvec{k})|_{\substack{stat}} = f(u_2) = 0$.
Explicitely, the stationary points of~$J_4$ with $u_2$ are
\begin{equation}
\tvec{k} = (\cos \phi, \sin \phi, 0 )^{\rm{T}},
\;
\phi \in \realnums,
\;
\text{with}
\; 
J_4(\tvec{k}) = 0.
\end{equation}
They are known as the ``D-flat'' directions, since they have $V_D=0$.
For the MSSM, they prevent the stability assertion by the quartic terms
alone.
For the stability to be guaranteed by $V_2>0$ in these
directions, theorem~\ref{theorem-stability} gives as
condition, see~\eqref{eq-gstabexcept} and~\eqref{eq-gdef}, the inequality
\begin{equation}
g(u_2) - \abs{\tvec{\xi}_\perp(u_2)} \sqrt{f'(u_2)}
  = \xi_0 - \sqrt{\xi_1^2 + \xi_2^2} > 0.
\end{equation}
Inserting~\eqref{eq-mssmp2} we get
\begin{equation}
\label{eq-mssmstab}
\abs{m_3^2} < \abs{\mu}^2 +\frac{1}{2}(m_{H_1}^2+m_{H_2}^2).
\end{equation}
as the necessary and sufficient condition for the stability of the MSSM
potential in the sense of~\eqref{eq-j4j2stab}.

\begin{figure}
\includegraphics[width=\linewidth]{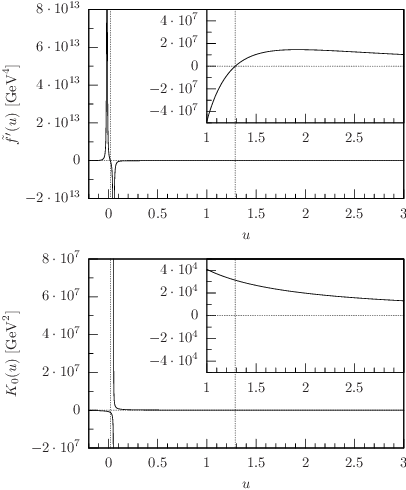}
\caption{\label{fig-mssmft}
The global minimum determining functions $\tilde{f}'(u)$ and $K_0(u)$
for the MSSM,
see~(\protect{\ref{eq-mssmftp}}) and~(\protect{\ref{eq-mssmk0}}),
with
$\abs{\mu}^2+m_{Hd}^2 = 157486\ \text{GeV}^2$,
$\abs{\mu}^2+m_{Hu}^2 = -2541\ \text{GeV}^2$,
$\abs{m_3^2} = 15341\ \text{GeV}^2$.
The small boxes show the functions with enhanced ordinate resolution
in the region around the physically relevant zero of $\tilde{f}'(u)$.
}
\end{figure}

For the global minimum to be non-trivial,
criterion~(I) of theorem~\ref{criteria-globalminima} gives
$\xi_0 < \sqrt{\xi_1^2+\xi_2^2+\xi_3^2}$, or equivalently
\begin{equation}
\label{eq-mssmtriv}
\abs{\mu}^2 +\frac{1}{2}(m_{H_1}^2 + m_{H_2}^2) < 
\sqrt{\abs{m_3^2}^2 + \frac{1}{4}\left( m_{H_1}^2 - m_{H_2}^2 \right)^2}
\end{equation}
as a necessary and sufficient condition.
We consider the acceptable global minimum candidates
given by the classes~(IIa) and (IIb) of theorem~\ref{classes-statpoints}.
The conditions~\eqref{eq-mssmstab} and \eqref{eq-mssmtriv}
prevent exceptional solutions. The regular solutions are
determined by the functions
\begin{align}
\label{eq-mssmft}
\tilde{f}(u) &= -\frac{1}{4}\left(
    \frac{\xi_0^2-\xi_1^2-\xi_2^2}{\frac{1}{8}g^2-u}
   -\frac{\xi_3^2}{-\frac{1}{8}g'^{\,2}-u}
  \right),\\
\label{eq-mssmftp}
\tilde{f}'(u) &= -\frac{1}{4}\left(
    \frac{\xi_0^2-\xi_1^2-\xi_2^2}{(\frac{1}{8}g^2-u)^2}
   -\frac{\xi_3^2}{(-\frac{1}{8}g'^{\,2}-u)^2}
 \right),\\
\label{eq-mssmk0}
K_0(u) &= -\frac{1}{2} \cdot
 \frac{\xi_0} {\frac{1}{8}g^2-u},
\end{align}
where we omitted the insertions~\eqref{eq-mssmp2}
for a compact notation.
Employing again the conditions~\eqref{eq-mssmstab} and \eqref{eq-mssmtriv}
we find the following.
The function $\tilde{f}'(u)$ always has two zeros and
those zeros imply values of $K_0(u)$ with opposite signs.
The physical solution with $K_0(u)>0$ has the Lagrange multiplier
\begin{equation}
u_0 = \frac{1}{8} \cdot
  \frac{ \abs{\xi_3} g^2 + \sqrt{\xi_0^2-\xi_1^2-\xi_2^2}\,g'^{\,2} }
       { \abs{\xi_3} - \sqrt{\xi_0^2-\xi_1^2-\xi_2^2} },
\end{equation}
which is positive.
Figure~\ref{fig-mssmft} shows the functions $\tilde{f}'(u), K_0(u)$
for an example set of parameters (corresponding to the
SPS1a scenario~\cite{Allanach:2002nj} at the tree level).
We conclude that \eqref{eq-mssmstab} and \eqref{eq-mssmtriv} guarantee
the existence of the stationary point $\fvec{K}(u_0)$,
which fulfils criterion~(II) of theorem~\ref{criteria-globalminima} and
therefore is the global minimum with the required EWSB pattern.
Moreover, there are no other local minima.

Note from~\eqref{eq-mssmp4} and \eqref{eq-mssmp2} that it is always
possible to choose a basis with \mbox{$\xi_1=-\abs{m_3^2}$},
$\xi_2=0$ without
affecting any other parameters.
We further choose a gauge where~\eqref{eq-aglo} and~\eqref{eq-gaugeneutral}
hold, and perform the rotation~\eqref{eq-goldrot} with
\begin{equation}
\tan\beta =
\sqrt{
  \frac{ \xi_0\abs{\xi_3} + \sqrt{\xi_0^2-\xi_1^2-\xi_2^2}\,\xi_3 }
       { \xi_0\abs{\xi_3} - \sqrt{\xi_0^2-\xi_1^2-\xi_2^2}\,\xi_3 }
}
\end{equation}
into a basis of the form~\eqref{eq-aglop} and \eqref{eq-atra},
which has the new parameters
\begin{gather}
\tvec{\xi}' = \begin{pmatrix}
  -c_{2\beta} \abs{m_3^2} - s_{2\beta}\, \frac{1}{2}(m_{H_1}^2-m_{H_2}^2) \\
  0\\
  -s_{2\beta} \abs{m_3^2} + c_{2\beta}\, \frac{1}{2}(m_{H_1}^2-m_{H_2}^2)
 \end{pmatrix},
\\[2mm]
E' = \frac{1}{8} \begin{pmatrix}
  -g^2 + s_{2\beta}^2\, \bar{g}^2 & 0 & 
     -\frac{1}{2} s_{4\beta}\, \bar{g}^2 \\
  0 & -g^2 & 0\\
  -\frac{1}{2} s_{4\beta}\, \bar{g}^2 & 0 &
     -g^2 + c_{2\beta}^2\, \bar{g}^2
  \end{pmatrix},
\end{gather}
with the abbreviations $\bar{g}^2:=g^2+g'^{\,2}$ and
$s_{2\beta}:=\sin 2\beta$ etc.
We insert the expressions into the formulae of Section~\ref{sec-potafter}
and use
\begin{equation}
m_W^2 := \left( \frac{1}{2}g v_0 \right)^2,
\qquad
m_Z^2 := \left( \frac{1}{2}\bar{g} v_0 \right)^2,
\end{equation}
with $v_0 = \sqrt{2 K_0(u_0)}$.
According to~\eqref{eq-mneu}, the fact that $\xi_2'=\eta_{12}'=0$
implies tree-level \CP conservation within the Higgs
sector of the MSSM.
We obtain the mass squares
\begin{equation}
m_{A^0}^2 = 2 v_0^2 (\eta_{22}' + u_0 ),
\quad
m_{H^\pm}^2 = m_{A^0}^2 + m_W^2
\end{equation}
for the pseudo-scalar boson $A^0:=h^{\prime \prime}$
and the charged bosons $H^\pm$, which are already mass eigenstates.
The non-diagonal part of the neutral mass matrix is
\begin{equation}
\left.{\cal M}^2\right|_{\begin{subarray}{l} \text{neutral,}\\
                               CP\text{ even} \end{subarray}} =
\begin{pmatrix}
c_{2\beta}^2\,m_Z^2 & -\frac{1}{2} s_{4\beta}\, m_Z^2 \\
-\frac{1}{2} s_{4\beta}\, m_Z^2 & m_{A^0}^2 + s_{2\beta}^2 m_Z^2
\end{pmatrix}
\end{equation}
in the basis $(\rho, h')$.
Its diagonalisation leads to mass squares
\begin{multline}
m_{h^0,H^0}^2 = 
   \frac{1}{2}\bigg(
   m_{A^0}^2+m_Z^2 \\
  \mp \sqrt{ \left(m_{A^0}^2+m_Z^2\right)^2
             - 4\,c_{2\beta}^2\,m_{A^0}^2 m_Z^2 }
  \bigg)
\end{multline}
for the mass eigenstates $h^0,H^0$.
They are obtained from
\begin{equation}
\label{eq-alpharot}
\begin{pmatrix} H^0 \\ h^0 \end{pmatrix}
=
\begin{pmatrix}\phantom{-}\cos \alpha' & \sin \alpha' \\
               -\sin \alpha' & \cos \alpha' \end{pmatrix}
\begin{pmatrix} \rho \\ h' \end{pmatrix},
\end{equation}
with the mixing angle $\alpha'$ determined by
\begin{equation}
\begin{split}
\cos 2\alpha' =&
  -\frac{m_{A^0}^2 - c_{4\beta}\, m_Z^2}{m_{H^0}^2-m_{h^0}^2},\\
\sin 2\alpha' =&
  -\frac{s_{4\beta}\, m_Z^2}{m_{H^0}^2-m_{h^0}^2}.
\end{split}
\end{equation}
Performing the rotation~\eqref{eq-goldrot} on the complete doublets,
as described above, leads to states $\rho,h'$
with simple couplings to the gauge bosons,
e.g. vanishing $ZZh'$ and $WWh'$ couplings at tree level.
However, usually the real parts of the neutral doublet components
are excluded from that rotation.
Applying the inverse of the rotation~\eqref{eq-goldrot} to only the
neutral components $(\rho, h')$ gives $\sqrt{2}(\mRe H_1^1, \mRe H_2^2)$.
The mass matrix in this basis is diagonalised analogously
to~\eqref{eq-alpharot}, where $\alpha'$ is replaced
by the mixing angle $\alpha$ with
\begin{equation}
\begin{split}
\cos 2\alpha =&
  -\cos{2\beta}\, \frac{m_{A^0}^2-m_Z^2}{m_{H^0}^2-m_{h^0}^2},\\
\sin 2\alpha =&
  -\sin{2\beta}\, \frac{m_{A^0}^2+m_Z^2}{m_{H^0}^2-m_{h^0}^2},
\end{split}
\end{equation}
which is the well-known result.


\subsection{Stationary points for THDM of Gunion et al.}
\label{ssec-firstex}

We continue the discussion of the potential
from~\cite{Gunion:1989we}, for which we derived the stability
conditions in Section~\ref{sec-stab}, see~\eqref{eq-gunp} et seq.
Note, that we consider the shifted potential according to $V(\fvec{K}=0)=0$.
The parameters of $V_4$ are given
by $\eta_{00}$, $\tvec{\eta}$ and $E$ as in~(\ref{eq-pagu}),
while we have
\begin{equation}
\begin{split}
\xi_0 &= -\lambda_1 v_1^2 - \lambda_2 v_2^2
        -2 \lambda_3 \big(v_1^2 + v_2^2\big),\\
\tvec{\xi} &= \begin{pmatrix}
    -v_1 v_2 \left( \lambda_5 \cos\xi
                    + \frac{\lambda_7}{2} \sin\xi \right)\\[2mm]
    -v_1 v_2 \left( \lambda_6 \sin\xi
                    + \frac{\lambda_7}{2} \cos\xi \right)\\[2mm]
    -\lambda_1 v_1^2 + \lambda_2 v_2^2
  \end{pmatrix}
\end{split}
\end{equation}
for the parameters of $V_2$.
Here, $v_1,v_2$ and $\xi$ denote the
parameters of the potential~(\ref{eq-gunp})
irrespective of their meaning for the vacuum expectation values.
The function $\tilde{f}'(u)$ constructed for this potential
according to~\eqref{eq-ftilpr} exhibits the zero
\begin{equation}
\tilde{\mu}=\frac{1}{4} \lambda_4,
\end{equation}
whose associated solution $\fvec{K}(\tilde{\mu})$
is always a stationary point.
It can be represented by the field configuration
\begin{equation}
\label{eq-gunobvious}
\left.\varphi_1\right|_{\substack{\tilde{\mu}}}
  = \begin{pmatrix} 0 \\ v_1 \end{pmatrix},
\qquad
\left.\varphi_2\right|_{\substack{\tilde{\mu}}}
  = \begin{pmatrix} 0 \\ v_2 \, e^{i \xi} \end{pmatrix}.
\end{equation}
If all $\lambda_i > 0$ it is immediately obvious from the definition
of the potential that this is the global minimum, which is furthermore
non-trivial unless both $v_1, v_2$ are zero.

However, the quartic stability conditions~\eqref{eq-gunstab}
do not require the $\lambda_i$ to be positive.
In the general case there can be more than one local minimum with
the required EWSB, and the stationary point~\eqref{eq-gunobvious} is not
necessarily the global minimum (we find parameters
where \eqref{eq-gunobvious} is only a saddle point, while
another solution provides an admissible global minimum for a stable
potential).
The function $\tilde{f}'(u)$ may have up to 5 additional zeros
which can lead to further regular stationary points.
Also exceptional stationary points may occur for special parameter
combinations.
We do not find analytical expressions for the remaining zeros
of $\tilde{f}'(u)$ for the general case.
Instead we apply the methods described in the previous sections
in a semi-analytical way and determine the zeros of $\tilde{f}'(u)$
numerically.
We assure the stability conditions~\eqref{eq-gunstab} to hold for
the chosen parameter values.
The stationary points with $K_0=\abs{\tvec{K}}>0$ and
the largest Lagrange multiplier $u>0$ are the required global
minima.
In order to classify also the other solutions, we compute $m_{H^\pm}^2$
and the eigenvalues of $\mathcal{M}_{\text{neutral}}^2$ for each
stationary point, as described in the previous sections
for the global minimum.
A solution for which all of these values are positive is a local minimum.
Mixed positive and negative values mean that the solution is a saddle
point.
Note that for a global minimum different from~\eqref{eq-gunobvious}, the
potential parameters $v_1,v_2$ and $\xi$ lose their simple meaning.
For example, we find that the global minimum acquires a
non-vanishing, \CP~violating, phase for certain parameters with
$\xi=0, \lambda_5=\lambda_6 \neq 0$ and $\lambda_7 \neq 0$.

\begin{figure*}
\includegraphics{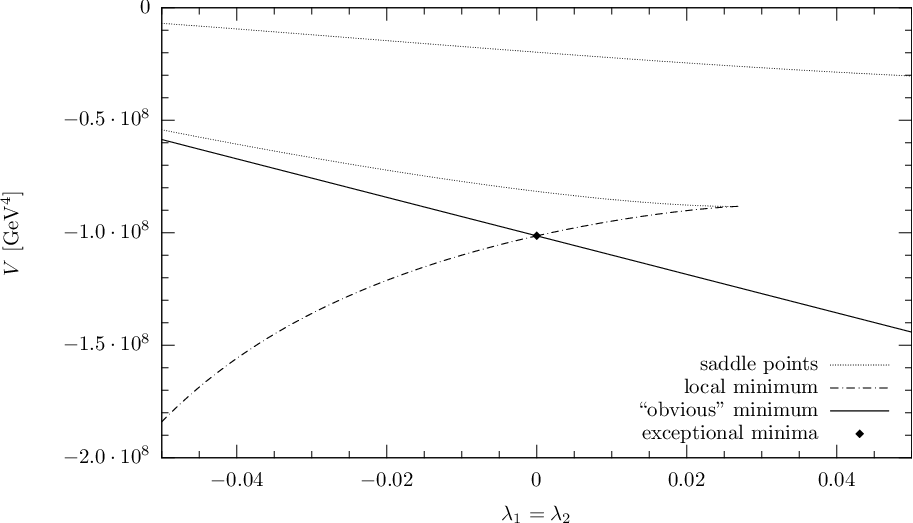}
\caption{\label{fig-vhunter}
The potential $V$ of~\cite{Gunion:1989we}, shifted to $V(\fvec{K}=0)=0$,
at all stationary points with $K_0 = \abs{\tvec{K}} > 0$
in dependence of $\lambda_1$, where $\lambda_2 = \lambda_1$.
The other parameters are
$\lambda_3=0.1,\ \lambda_4=0.2$, \mbox{$\lambda_5=\lambda_6=0.4$},
$v_1=30\text{~GeV},\ v_2=171\text{~GeV},\ \lambda_7=0,\ \xi=0$.
The lines represent regular stationary points,
where the solid curve corresponds to the ``obvious'' solution
with $\tan\beta=v_2/v_1$ and $v_0=\sqrt{2(v_1^2+v_2^2)}$,
which is a local minimum for the chosen parameters.
For $\lambda_1=\lambda_2=0$ there are two degenerate
exceptional minima
and regular solutions only for the saddle points.
Depending on $\lambda_1=\lambda_2$,
the global minimum is given by that local minimum out of the two
which has the lower potential.
}
\end{figure*}

Figure~\ref{fig-vhunter} shows the potential
at the stationary points with $K_0=\abs{\tvec{K}}>0$
for the particular parameter values described in the caption.
This example features tree-level \CP conservation
within the Higgs sector:
\mbox{$\xi=\lambda_7=0$} leads to $\xi_2=0$ and, by~\eqref{eq-solktil}
and $\lambda_4 \neq \lambda_6$,
to $K_2=0$ (i.e. trivial phases) at the global minimum.
In the basis with~\eqref{eq-aglop} and~\eqref{eq-atra}
we then have $\xi_2'=\eta_{12}'=0$,
implying \CP conservation by~\eqref{eq-mneu}.
Note that even with the simple parameter combinations chosen for
Fig.~\ref{fig-vhunter},
the structure of the stationary points can be non-trivial.
In the example $\lambda_1$ and $\lambda_2$ are equal and
varied simultaneously.
For the plotted range, where $\lambda_1$ is negative, the global minimum
is a regular solution and differs from~\eqref{eq-gunobvious},
which is only a local minimum.
For $\lambda_1=0$ there are two exceptional and degenerate minima.
The figure shows for positive $\lambda_1$ the expected behaviour,
that~\eqref{eq-gunobvious} becomes the global minimum, but also that for
$0<\lambda_1< 0.0268$ a second regular local minimum
exists.
Two stationary points disappear for the plotted range
above $\lambda_1 > 0.0268$ because there the corresponding two zeros of
$\tilde{f}'(u)$ have a non-vanishing imaginary part.

\section{Conclusions}
\label{sec-conclu}

We have analysed the scalar potential of the general THDM.
In order to give an acceptable theory, this potential has to obey
certain criteria.
The potential should be stable, that is bounded from below, and lead
to the EWSB pattern observed in Nature.
The conditions found for the stability of the potential and
for~EWSB are transparent and compact if the potential is written down
in terms of gauge invariant field functions~\eqref{eq-kmatdecomp}.
These conditions allow for a simple application to every specific
THDM. We illustrated our method in two examples, namely the 
MSSM potential as well as the THDM potential introduced by
Gunion et al.~\cite{Gunion:1989we}.
In the case of the MSSM we could easily reproduce all well-known results. 
For the potential of Gunion et al. we could clarify some interesting 
aspects of the model, see Figure~\ref{fig-vhunter}.

We note that the method presented here may also be extendable to general
Multi-Higgs-Doublet Models.
A first step in this direction is done in Appendix~\ref{app-ndoublets}.
Further, in a more detailed study it is mandatory to take
quantum corrections to the Higgs potential into account.
For the resulting effective Higgs potential the conditions for stability and
for~EWSB are then in general modified.

\begin{acknowledgments}
We thank R.~Barbieri, A.~Brandenburg, Q.~Shafi, W.~Wetzel and P.M.~Zerwas
for valuable discussions.
This work was supported by the German Bundesministerium f\"ur Bildung
und Forschung, project numbers HD05HT1VHA/0 and HD05HT4VHA/0.
\end{acknowledgments}


\appendix
\section{Gauge orbits}
\label{app-orbits}

Here we give the proof that the gauge orbits of the Higgs fields of the THDM
are parametrised by four vectors $(K_0,\tvec{K})$ satisfying~\eqref{eq-kinq},
or equivalently, by positive-semidefinite matrices
$\twomat{K}$~\eqref{eq-kmat},~\eqref{eq-kmatdecomp}.
Indeed, let us consider two Higgs doublets as in~\eqref{eq-doubldef},
\begin{equation}
\varphi_i^\alpha(x),\quad
i=1,2,\quad
\alpha=+,0.
\end{equation}
We arrange these in a \mbox{$2 \by 2$}~matrix
\begin{equation}
\label{eq-phimatdef}
\phi(x) :=
  \left( \varphi_i^{ \alpha} (x) \right) =
  \begin{pmatrix} \varphi_1^+(x) & \varphi_1^0(x) \\
                  \varphi_2^+(x) & \varphi_2^0(x) \end{pmatrix}.
\end{equation}
Then we have from~\eqref{eq-kmat}
\begin{equation}
\label{eq-kmatphi}
\twomat{K}(x) = \phi(x) \phi^\dagger(x).
\end{equation}
The change of basis~\eqref{eq-udef} means the transformation
\begin{equation}
\phi(x) \rightarrow \phi'(x) = U \phi(x).
\end{equation}
A gauge transformation from the \eweakgroup gauge group means
\begin{equation}
\varphi_i^\alpha(x) \rightarrow \varphi_i'^\alpha(x)
  = \left(U_G(x)\right)_{\alpha\beta} \varphi_i^\beta(x),
\end{equation}
where
\begin{equation}
U_G(x) \in U(2).
\end{equation}
Thus, under a gauge transformation the matrix $\phi(x)$ behaves as
\begin{equation}
\label{eq-phigtrafo}
\phi(x) \rightarrow \phi'(x) = \phi(x) U_G^\trans(x).
\end{equation}

As we discussed in Section~\ref{sec-thdm} any matrix $\twomat{K}(x)$ formed
from the Higgs fields according to~\eqref{eq-kmat},
which is equivalent to~\eqref{eq-kmatphi}, must be positive semidefinite.
Conversely, given any positive semidefinite matrix~$\twomat{K}(x)$ we
can diagonalise it by a \mbox{$2 \by 2$}~unitary transformation $W(x)$:
\begin{equation}
\begin{split}
\label{eq-kmatdiagw}
\twomat{K}(x)
  =& W(x) \begin{pmatrix} \kappa_1(x) & 0 \\ 0 & \kappa_2(x) \end{pmatrix} W^\dagger(x),\\
&W^\dagger(x) W(x) = \unitmatrix .
\end{split}
\end{equation}
Since we have $\kappa_1(x)\geq 0$ and $\kappa_2(x)\geq 0$ we can set
\begin{equation}
\phi(x)
  = W(x) \begin{pmatrix} \sqrt{\kappa_1(x)} & 0 \\ 0 & \sqrt{\kappa_2(x)} \end{pmatrix}
\end{equation}
and get
\begin{equation}
\twomat{K}(x) = \phi(x) \phi^\dagger(x).
\end{equation}
With this we have proven the following.
\begin{itemize}
\item
For any positive semidefinite matrix $\twomat{K}(x)$ there are Higgs fields
satisfying~\eqref{eq-kmatphi} respectively~\eqref{eq-kmat}.
\end{itemize}

Now suppose that we have a given positive semidefinite matrix~$\twomat{K}(x)$
and two Higgs-field matrices~$\phi(x),\phi'(x)$, both
satisfying~\eqref{eq-kmatphi},
\begin{equation}
\label{eq-samekmat}
\twomat{K}(x) = \phi(x)\phi^\dagger(x) = \phi'(x) \phi'^\dagger(x).
\end{equation}
We want to show that $\phi'(x)$ and $\phi(x)$ are then related by a gauge
transformation~\eqref{eq-phigtrafo}.
We consider three cases.
\begin{enumerate}
\item
$\twomat{K}(x)=0$. Then $\phi(x)=\phi'(x)=0$ and~\eqref{eq-phigtrafo}
is trivially fulfilled.
\item
$\twomat{K}(x)>0$, that is $\twomat{K}(x)$ is positive definite. Then
\begin{equation}
\label{eq-regularphi}
\det \twomat{K}(x) = \abs{\det\phi(x)}^2 = \abs{\det\phi'(x)}^2 >0
\end{equation}
and both $\phi(x)$ and $\phi'(x)$ have an inverse. We set
\begin{equation}
\phi^{-1}(x)\phi'(x) = U_G^\trans(x)
\end{equation}
and find from~\eqref{eq-samekmat}
\begin{equation}
U_G^\dagger(x) U_G(x) = \unitmatrix,
\end{equation}
that is $U_G(x) \in U(2)$, and
\begin{equation}
\phi'(x) = \phi(x) U_G^\trans(x).
\end{equation}
Thus $\phi'(x)$ and $\phi(x)$ satisfy~\eqref{eq-phigtrafo}, they
are related by a gauge transformation.
\item
$\twomat{K}(x)$ has rank 1, that is the eigenvalues are
\begin{equation}
\label{eq-kmatr1evals}
\kappa_1(x) > 0, \qquad \kappa_2(x)=0 .
\end{equation}
With the matrix~$W(x)$ diagonalising~$\twomat{K}(x)$ (see~\eqref{eq-kmatdiagw})
we have then from~\eqref{eq-samekmat}
\begin{equation}
\begin{split}
\qquad\begin{pmatrix} \kappa_1(x) & 0 \\ 0 & 0 \end{pmatrix}
 &= \left( W^\dagger(x)\phi(x) \right)
   \left( W^\dagger(x)\phi(x) \right)^\dagger\\
 &= \left( W^\dagger(x)\phi'(x) \right)
   \left( W^\dagger(x)\phi'(x) \right)^\dagger .
\end{split}
\end{equation}
From this we see that
\begin{equation}
\begin{split}
W^\dagger(x)\phi(x)
  &= \begin{pmatrix} \chi_1^+(x) & \chi_1^0(x) \\ 0 & 0 \end{pmatrix},\\
W^\dagger(x)\phi'(x)
  &= \begin{pmatrix} \chi_1'^{\,+}(x) & \chi_1'^{\,0}(x) \\ 0 & 0 \end{pmatrix},
\end{split}
\end{equation}
where
\begin{equation}
\chi_1^\dagger(x)\chi_1(x) = \chi_1'^{\,\dagger}(x) \chi_1'(x) = \kappa_1(x),
\end{equation}
with the composition of the vectors~$\chi_i(x), \chi_i'(x)$
defined as for the vectors~$\varphi_i(x)$.
Therefore we can find a matrix $U_G(x)\in U(2)$ such that
\begin{equation}
\chi_1'^{\,\alpha}(x) = \left( U_G(x) \right)_{\alpha\beta} \chi_1^\beta(x),
\end{equation}
which implies
\begin{equation}
\begin{split}
W(x) \phi'(x) &= W(x) \phi(x) U_G^\trans(x),\\
\phi'(x) &= \phi(x) U_G^\trans(x).
\end{split}
\end{equation}
That is $\phi'(x)$ and $\phi(x)$ are related by a gauge transformation.
\end{enumerate}
With this we have completed the proof of the following statement.

\begin{theorem}
\begin{samepage}
Any two Higgs-doublet fields giving the same
matrix~$\twomat{K}(x)$~\eqref{eq-kmat}, respectively~\eqref{eq-kmatphi},
are related by a gauge transformation.
The space of gauge orbits can be parametrised by four-vectors~$(K_0,\tvec{K})$
lying on and inside the forward light cone, see~\eqref{eq-kinq}.
\end{samepage}
\end{theorem}

We close this appendix with a supplementary note to Section~\ref{sec-ssbr}
on the breaking of the \eweakgroup gauge group.
The matrix~(\ref{eq-phimatdef}) of vacuum expectation values is
\begin{equation}
\phi_{vac} = \begin{pmatrix} v_1^+ & v_1^0 \\ v_2^+ & v_2^0 \end{pmatrix} .
\end{equation}
If the global minimum of the potential occurs with $K_0>\abs{\tvec{K}}$
the corresponding matrix~$\twomat{K}$ has rank~2.
Then from~\eqref{eq-kmatphi} and \eqref{eq-regularphi} we see that also
$\phi_{vac}$ has rank 2.
Invariance of $\phi_{vac}$ under a gauge
transformation~\eqref{eq-phigtrafo}
\begin{equation}
\label{eq-phivacfull}
\phi_{vac} = \phi_{vac}\, U_G^\trans
\end{equation}
is then only possible for $U_G= \unitmatrix$.
That is how we see with the methods of this appendix that in
this case the full gauge group \eweakgroup is broken.


\section{The case of \mbox{$n$ doublets}}
\label{app-ndoublets}

In this appendix we generalise the methods of Section~\ref{sec-thdm} and
Appendix~\ref{app-orbits} to the case of \mbox{$n>2$} Higgs doublets.
We consider~$n$ complex Higgs-doublet fields
\begin{equation}
\label{eq-ndoubldef}
\varphi_i(x) = \begin{pmatrix} \varphi^+_i(x) \\  \varphi^0_i(x) \end{pmatrix},
\quad
i=1,\ldots,n .
\end{equation}
All doublets are supposed to have the same weak hypercharge \mbox{$y=+1/2$}.
In analogy to~(\ref{eq-kmat}) we introduce the matrix
\begin{equation}
\label{eq-nkmat}
\twomat{K}(x) =
\left( \twomat{K}_{ij}(x) \right) :=
\left( \varphi_j^{\dagger}(x) \varphi_i(x) \right),
\end{equation}
which is now a \mbox{$n \by n$} matrix.
The aim is to discuss the properties of $\twomat{K}(x)$.
For this we introduce the \mbox{$n \by n$} matrix~$\phi(x)$
(compare~(\ref{eq-phimatdef}))
\begin{equation}
\label{eq-nphimat}
\phi(x) :=
\begin{pmatrix}
  \varphi_1^{+}(x) & \varphi_1^0(x) & 0 & \ldots & 0 \\
  \varphi_2^{+}(x) & \varphi_2^0(x) & 0 & \ldots & 0 \\
  \vdots      & \vdots   & \vdots & \cdots & \vdots\\
  \varphi_n^{+}(x) & \varphi_n^0(x) & 0 & \ldots & 0
\end{pmatrix}.
\end{equation}
It is easy to see that we have
\begin{equation}
\label{eq-nkdec}
\twomat{K}(x) = \phi(x) \, \phi^\dagger(x).
\end{equation}
A change of basis among the doublets means
\begin{equation}
\label{eq-nchbasis}
\phi(x) \rightarrow \phi'(x) = U \, \phi(x)
\end{equation}
with a constant matrix~\mbox{$U \in U(n)$},
\begin{equation}
\label{eq-nunitary}
U^{\dagger} \, U = \unitmatrix_n.
\end{equation}
A gauge transformation from \eweakgroup 
means
\begin{equation}
\label{eq-ngaugetrans}
\phi(x) \rightarrow \phi'(x)= \phi(x) \, \tilde{U}_G^\trans(x),
\end{equation}
where $\tilde{U}_G(x)$ is block-diagonal
\begin{equation}
\label{eq-UGtrans}
\tilde{U}_G(x) :=
\begin{pmatrix}
  U_G(x) &\vline& 0 \\
  \hline
  0             &\vline & \unitmatrix_{n-2}
\end{pmatrix},
\end{equation}
\mbox{$U_G(x) \in U(2)$} and thus \mbox{$\tilde{U}_G(x) \in U(n)$}. 
We have then from (\ref{eq-ngaugetrans})
\begin{equation}
\label{eq-nphiprime}
\varphi_i'^{\,\alpha}(x) = \left( U_G(x) \right)_{\alpha \beta} \, \varphi_i^{\,\beta}(x),
\qquad i=1,\ldots,n.
\end{equation}
From~(\ref{eq-nphimat}) and~(\ref{eq-nkdec}) we see
that the matrix~$\twomat{K}(x)$ has the following properties
\begin{itemize}
\item{$\twomat{K}(x)$ is positive semidefinite,}
\item{$\twomat{K}(x)$ has rank \mbox{$\le 2$}.}
\end{itemize}
That is, $\twomat{K}(x)$ has at most two
eigenvalues \mbox{$\kappa_1(x), \kappa_2(x)>0$} and the
remaining eigenvalues \mbox{$\kappa_3(x), \ldots, \kappa_n(x)$} must be zero.
The rank condition can be seen as follows.
We denote by \mbox{$\psi^+(x)$, $\psi^0(x)$}
the first two column vectors of $\phi(x)$.
Then we have
\begin{align}
\label{eq-nphivec}
\phi(x) &= \bigg( \psi^+(x),  \psi^0(x),  0,  \ldots,  0 \bigg),\\
\nonumber
\twomat{K}(x) &=
\bigg(
\varphi_1^{+ *}(x)\, \psi^+(x) + \varphi_1^{0 *}(x)\, \psi^0(x), 
\ldots,\\
\label{eq-nKvec}
& \qquad \varphi_n^{+ *}(x)\, \psi^+(x) + \varphi_n^{0 *}(x)\, \psi^0(x)
\bigg).
\end{align}
That is, at most two column vectors of $\twomat{K}(x)$
are linearly independent.

Suppose now that we have a given positive semidefinite matrix
$\twomat{K}(x)$ of rank \mbox{$\le 2$}. Then we can
diagonalise $\twomat{K}(x)$ and represent it as
\begin{equation}
\label{eq-nKmatdiag}
\twomat{K}(x) =
W(x)
\begin{pmatrix}
  \begin{matrix}
    \kappa_1(x) & 0\\
    0      & \kappa_2(x)
  \end{matrix}
  & \vline & 0\\
  \hline
  0 & \vline & 0
\end{pmatrix}
W^{\dagger}(x)
\end{equation}
with \mbox{$W(x) \in U(n)$} and \mbox{$\kappa_1(x) \ge 0$},
\mbox{$\kappa_2(x) \ge 0$}.
We set now
\begin{equation}
\label{eq-nphidiag}
\phi(x) =
W(x)
\begin{pmatrix}
  \begin{matrix}
    \sqrt{\kappa_1(x)} & 0\\
    0             & \sqrt{\kappa_2(x)}
  \end{matrix}
  & \vline & 0\\
  \hline
  0 &\vline & 0
\end{pmatrix}
\end{equation}
and see easily that $\phi(x)$ is of the form~(\ref{eq-nphimat})
and satisfies~(\ref{eq-nkdec}). Thus to any positive
semidefinite matrix $\twomat{K}(x)$ of rank~$\le 2$ there is
at least one field configuration of the~$n$ Higgs doublets
such that~(\ref{eq-nkmat}) holds.

Suppose now that we have two field configurations, that is two
matrices~$\phi(x)$ and $\phi'(x)$ of the type~(\ref{eq-nphimat})
such that
\begin{equation}
\label{eq-nkdec2}
\twomat{K}(x) = \phi(x)\, \phi^\dagger(x)=\phi'(x)\, \phi'^{\,\dagger}(x).
\end{equation}
We can diagonalise~$\twomat{K}(x)$ as in~(\ref{eq-nKmatdiag})
and get
\begin{equation}
\begin{split}
\label{eq-nKmatcalc}
\begin{pmatrix}
  \begin{matrix}
    \kappa_1(x) & 0\\
    0      & \kappa_2(x)
  \end{matrix}
  & \vline & 0\\
  \hline
  0 & \vline & 0
\end{pmatrix}
=&
\left( W^{\dagger}(x)\, \phi(x) \right)
\left( W^{\dagger}(x)\, \phi(x) \right)^{\dagger}\\
=&
\left( W^{\dagger}(x)\, \phi'(x) \right)
\left( W^{\dagger}(x)\, \phi'(x) \right)^{\dagger}.
\end{split}
\end{equation}
From this we see that we must have
\begin{align}
W^{\dagger}(x) \, \phi(x) &=
\begin{pmatrix}
  \begin{matrix}
    \,\chi_1^+(x)\, & \,\chi_1^0(x)\,\\
    \,\chi_2^+(x)\, & \,\chi_2^0(x)\,
  \end{matrix}
  & \vline & 0\\
  \hline
  0 & \vline & 0
\end{pmatrix},\\
W^{\dagger}(x) \, \phi'(x) &=
\begin{pmatrix}
  \begin{matrix}
    \chi_1'^{\,+}(x) & \chi_1'^{\,0}(x)\\
    \chi_2'^{\,+}(x) & \chi_2'^{\,0}(x)
  \end{matrix}
  & \vline & 0\\
  \hline
  0 & \vline & 0
\end{pmatrix},
\end{align}
where
\begin{align}
\begin{split}
\chi_1^{\dagger}(x) \, \chi_1(x) &= \chi_1'^{\,\dagger}(x) \, \chi_1'(x) = \kappa_1(x),\\
\chi_2^{\dagger}(x) \, \chi_2(x) &= \chi_2'^{\,\dagger}(x) \, \chi_2'(x) = \kappa_2(x),\\
\chi_1^{\dagger}(x) \, \chi_2(x) &= \chi_1'^{\,\dagger}(x) \, \chi_2'(x) = 0.
\end{split}
\end{align}
From this we conclude that we can find a matrix~\mbox{$U_G(x) \in U(2)$}
such that
\begin{equation}
\chi_i'^{\,\alpha} =
\left( U_G(x) \right)_{\alpha \beta} \, \chi^{\beta}_i(x), \qquad i=1,2.
\end{equation}
Inserting this $U_G(x)$ into~(\ref{eq-UGtrans}) we get
\begin{equation}
W^{\dagger}(x) \phi'(x) =
W^{\dagger}(x) \phi(x) \, \tilde{U}_G^T(x)
\end{equation}
and since~\mbox{$W(x) \in U(n)$}
\begin{equation}
\phi'(x) =
\phi(x) \tilde{U}_G^T(x).
\end{equation}
That is $\phi'(x)$ and $\phi(x)$ are related by a gauge transformation.
We summarise our findings in a theorem.

\begin{theorem}
\begin{samepage}
For $n$ Higgs-doublet fields of the same weak hypercharge~$y=+1/2$
the matrix
$
\twomat{K}(x) =
\left( \varphi_j^{\dagger}(x) \varphi_i(x) \right)
$
is a positive semidefinite
\mbox{$n \by n$}~matrix of rank~$\le 2$.
For any positive semidefinite \mbox{$n \by n$}~matrix~$\twomat{K}(x)$
of rank~$\le 2$ there are Higgs fields such that~(\ref{eq-nkmat})
holds. Any two field configurations giving the same matrix
$\twomat{K}(x)$ are related by a \eweakgroup gauge
transformation. The matrices~$\twomat{K}(x)$ form, therefore,
the space of the gauge orbits of the~$n$ Higgs-doublet fields.
\end{samepage}
\end{theorem}

As an example we consider three Higgs doublets. The space of
gauge orbits is then given by all positive semidefinite~\mbox{$3 \by 3$}
matrices~$\twomat{K}(x)$ with~\mbox{$\det \twomat{K}(x)=0$}.


\end{document}